%% file: ms.tex
\documentclass{emulateapj}

\def\hda{\mbox{HD~108317}}
\def\hdb{\mbox{HD~122563}}
\def\hdc{\mbox{HD~126238}}
\def\hdd{\mbox{HD~128279}}
\def\bd{\mbox{BD~$+$17~3248}}

\def\kmsec{\mbox{km~s$^{\rm -1}$}}
\def\logg{\mbox{log~{\it g}}}

\def\teff{\mbox{$T_{\rm eff}$}}
\def\vt{\mbox{$v_{\rm t}$}}
\def\rpro{\mbox{$r$-process}}
\def\spro{\mbox{$s$-process}}

\def\loggf{$\log$($gf$)}
\def\loggfalt{$\log gf$}

\shorttitle{New NUV $n$-capture Abundances}
\shortauthors{Roederer et al.}
\slugcomment{Accepted for publication in the Astrophysical Journal Supplement
Series}

\begin{document}

\title{
New \textit{Hubble Space Telescope} Observations of Heavy Elements \\
in Four Metal-Poor 
Stars\footnotemark[1]}

\footnotetext[1]{
Based on observations made with the NASA/ESA Hubble Space Telescope, 
obtained at the Space Telescope Science Institute, 
which is operated by the Association of Universities for Research in 
Astronomy, Inc., under NASA contract NAS~5-26555. 
These observations are associated with programs 8111 and 12268. 
This paper includes data gathered with the 6.5~meter 
Magellan Telescopes located at Las Campanas Observatory, Chile. 
Some of the data presented herein were obtained at the W.M.\ Keck Observatory, 
which is operated as a scientific partnership among the 
California Institute of Technology, the University of California, and 
the National Aeronautics and Space Administration. 
The Observatory was made possible by the generous financial support of the 
W.M.\ Keck Foundation.
}

\author{
Ian U.\ Roederer,\altaffilmark{2}
James E.\ Lawler,\altaffilmark{3}
Jennifer S.\ Sobeck,\altaffilmark{4}
Timothy C.\ Beers,\altaffilmark{5,}\altaffilmark{6,}\altaffilmark{7}
John J.\ Cowan,\altaffilmark{8} \\
Anna Frebel,\altaffilmark{9}
Inese I.\ Ivans,\altaffilmark{10}
Hendrik Schatz,\altaffilmark{6,}\altaffilmark{7,}\altaffilmark{11}
Christopher Sneden,\altaffilmark{12}
Ian B.\ Thompson\altaffilmark{2}
}

\altaffiltext{2}{Carnegie Observatories, 
Pasadena, CA 91101, USA
}
\altaffiltext{3}{Department of Physics, University of Wisconsin, 
Madison, WI 53706, USA
}
\altaffiltext{4}{Department of Astronomy \& Astrophysics, 
University of Chicago, Chicago, IL 60637, USA
}
\altaffiltext{5}{National Optical Astronomy Observatory, Tucson, AZ 85719,
USA
}
\altaffiltext{6}{Department of Physics \& Astronomy,
Michigan State University, E.\ Lansing, MI 48824, USA
}
\altaffiltext{7}{Joint Institute for Nuclear Astrophysics, 
Michigan State University, E.\ Lansing, MI  48824, USA
}
\altaffiltext{8}{Homer L.\ Dodge Department of Physics and Astronomy,
University of Oklahoma, 
Norman, OK 73019, USA
}
\altaffiltext{9}{Massachusetts Institute of Technology, 
Kavli Institute for Astrophysics and Space Research, 
Cambridge, MA 02139, USA
}
\altaffiltext{10}{Department of Physics and Astronomy, University of Utah,
Salt Lake City, UT 84112, USA
}
\altaffiltext{11}{National Superconducting Cyclotron Laboratory, 
Michigan State University, East Lansing, MI 48824, USA
}
\altaffiltext{12}{Department of Astronomy, University of Texas at Austin,
Austin, TX 78712, USA
}


\addtocounter{footnote}{12}

\begin{abstract}

Elements heavier than the iron group are found in nearly
all halo stars.
A substantial number of these elements, key to understanding
neutron-capture nucleosynthesis mechanisms,
can only be detected in the near-ultraviolet.
We report the results of an observing campaign using the
Space Telescope Imaging Spectrograph on board the
\textit{Hubble Space Telescope} to study the detailed
heavy element abundance patterns in four metal-poor stars.
We derive abundances or upper limits from
27~absorption lines of 15~elements produced by neutron-capture reactions,
including seven elements (germanium, cadmium, tellurium, 
lutetium, osmium, platinum, and gold) that 
can only be detected in the
near-ultraviolet.
We also examine 202 heavy element absorption lines
in ground-based optical spectra obtained with the
Magellan Inamori Kyocera Echelle 
Spectrograph
on the Magellan-Clay Telescope at Las Campanas Observatory and
the High Resolution Echelle Spectrometer 
on the Keck~I Telescope on Mauna Kea.
We have detected up to 34~elements heavier than zinc.
The bulk of the heavy elements in these four stars 
are produced by 
$r$-process nucleosynthesis.
These observations affirm earlier results
suggesting that the tellurium
found in metal-poor halo stars with moderate amounts of
$r$-process material scales with the
rare earth and third $r$-process peak elements.
Cadmium often follows the abundances of the neighboring
elements palladium and silver.
We identify several sources of systematic uncertainty that 
must be considered when comparing these abundances with
theoretical predictions.
We also present new isotope shift and hyperfine structure
component patterns
for Lu~\textsc{ii} and Pb~\textsc{i} lines
of astrophysical interest.

\end{abstract}

\keywords{
nuclear reactions, nucleosynthesis, abundances---stars: abundances---stars: 
individual (HD~108317, HD~122563, HD~126238, HD~128279)---stars: Population II
}

\section{Introduction}
\label{intro}

Thousands of high-resolution spectroscopic observations 
acquired over the last few decades
have revealed that nearly all stars contain at least
traces of elements heavier than the iron group.
These elements are primarily produced by neutron-capture
reactions, and 
their star-to-star abundance variations span 
several orders of magnitude.
Two general reaction timescales are involved, those
that are slow (the \spro) or rapid (the \rpro) 
relative to the average $\beta$-decay times for unstable nuclei
along the reaction paths.
The abundance distributions are 
governed by nuclear structure
and the physical conditions 
at the time of nucleosynthesis.

The solar system (S.S.) composition represents one chemical
snapshot of the Galactic interstellar medium 
4.5~Gyr ago.
Models predict that about 51\% (by mass or by number) 
of the heavy elements in the S.S.\ originated
through \spro\ nucleosynthesis, with the
remaining 49\% produced by other nucleosynthesis mechanisms, primarily
the \rpro.
This is revealed by comparing the
S.S.\ isotopic abundance distribution with predictions
from analytical models or nuclear reaction networks
coupled to stellar evolution codes
(e.g., \citealt{seeger65,cameron73,cameron82,kappeler89,arlandini99,
bisterzo11}).
These models predict the \spro\ component, which is
normalized to isotopes that
can only be produced by \spro\ nucleosynthesis.
This distribution is subtracted from 
the total abundance distribution to reveal
the \rpro\ ``residuals.''
Small contributions from mechanisms that produce proton-rich nuclei
can be subtracted in a similar way, but
this is negligible for most elements.
The \rpro\ residuals implicitly include contributions from 
all other processes.

Strontium (Sr, $Z =$~38) or barium (Ba, $Z =$~56)
have been detected, or not excluded based on upper limits,
in nearly all metal-poor field and globular cluster stars studied
(e.g., \citealt{mcwilliam95,ryan96,honda04,aoki05,francois07,cohen08,lai08,
roederer10c,roederer11,roederer12d,hollek11}), including
the most iron-poor star known \citep{frebel05}.
These elements are also found in all dwarf galaxies
studied, even those with tight upper limits
on strontium or barium in a few stars
\citep{fulbright04,koch08,frebel10}.
This early and widespread enrichment, presumably by some form
of \rpro\ nucleosynthesis,
argues for a 
nucleosynthetic site associated with common, short-lived stars.
Models and simulations of the
neutrino wind of core-collapse supernovae 
can reproduce a number of observational details
(e.g., \citealt{mathews92,woosley94,wheeler98,wanajo03,farouqi10,peterson11}).
There are concerns, however, that conditions favorable to the 
``main'' component of the
\rpro---a neutron-rich wind, for example---may not actually develop
(e.g., \citealt{horowitz99,janka08,fischer10,hudepohl10,roberts10}).
Merging neutron stars or neutron star-black hole binaries
are attractive candidate sites also
(e.g., \citealt{freiburghaus99,goriely11}),
though they may have difficulty
producing prompt \rpro\ enrichment in nearly all stellar environments
(e.g., \citealt{argast04,wanajo06}).

Stars in the hydrogen and helium 
shell-burning stage on the asymptotic giant branch
are major sites of \spro\ nucleosynthesis
(e.g., \citealt{busso99}), 
and massive stars with high metallicity
may activate \spro\ reactions during the
core helium and shell carbon burning stages (e.g., \citealt{raiteri93}).
Many low metallicity halo giants enriched with 
substantial amounts of \spro\ material
are found in binary systems (e.g., \citealt{mcclure84,mcclure90}).
In contrast, radial-velocity monitoring of stars
with \rpro\ enhancements from $+0.5 \leq$~[Eu/Fe]~$\leq +$1.8
demonstrates that the binary fraction
of these stars is no different than other low-metallicity
halo stars (about 20\%; \citealt{hansen11b}).
This suggests that substantial levels of \rpro\ enrichment
do not require the presence of a binary companion,
underscoring the likelihood that the astrophysical
sites of the $s$- and \rpro\ are decoupled from one another.
Yet \spro\ models cannot reproduce all of the 
abundance characteristics of another class of stars
showing significant excesses of both $s$- and \rpro\ material
\citep{lugaro12}.
Studies like these highlight the need to identify the astrophysical
site or sites of \rpro\ nucleosynthesis.

The distributions of \rpro\ material
in the S.S.\ and halo stars provide
constraints on the physical conditions at these candidate sites.
While the S.S.\ distribution is of great value, it exhibits
several shortcomings.
The S.S.\ \rpro\ residual distribution is 
only as reliable as the S.S.\ total and
\spro\ distributions.
The most precise isotopic abundance measurements are derived from
the study of rare and fragile CI chondrite meteorites, 
which have experienced the least amount of
fractionation (e.g., \citealt{anders71}).
The S.S.\ \rpro\ distribution represents
the combined yields of many \rpro\ events,
so interesting but rare features may be averaged out.
The detailed \rpro\ patterns observed in
metal-poor stars offer an independent set of constraints,
but they present a different set of challenges.
Not all elements can be detected in stellar spectra, and
only elemental---not isotopic---abundances are
generally available.
Stellar abundances are derived, not measured, 
and require an accurate stellar atmosphere model and various atomic data.
Interpreting the derived abundance pattern as being
representative of only \rpro\ nucleosynthesis 
presents a final challenge.

Despite these challenges, the S.S.\ meteoritic and solar 
photospheric abundances agree remarkably well for most elements
(e.g., \citealt{anders89}).
The scaled S.S.\ \rpro\ distribution also
closely matches that
observed in metal-poor halo stars strongly enriched in the
\rpro, such as \mbox{CS~22892--052}
\citep{cowan95,sneden03}.
This similarity is of great interest, but so too are
the differences from one stellar \rpro\ distribution to another 
(e.g., \citealt{mcwilliam98,sneden00,hill02,
johnson02,aoki05,barklem05,ivans06,frebel07,honda07,roederer10c,
peterson11,hansen12}).

Some elements exhibit greater sensitivity
to different physical aspects of the \rpro,
so it is important to try to detect these
rarely-seen elements in stellar spectra.
The high-resolution near-ultraviolet (NUV) capabilities of the
Goddard High Resolution Spectrograph (GHRS) and
the Space Telescope Imaging Spectrograph (STIS)
on board the \textit{Hubble Space Telescope} (\textit{HST})
have been especially helpful in this regard.
Without them, 
many heavy elements would forever remain undetected---or with
marginal detections only---in 
\rpro\ enriched stars.
These include 
germanium (Ge, $Z =$~32),
arsenic (As, $Z =$~33),
selenium (Se, $Z =$~34),
molybdenum (Mo, $Z =$~42),
cadmium (Cd, $Z =$~48), 
tellurium (Te, $Z =$~52),
lutetium (Lu, $Z =$~71),
osmium (Os, $Z =$~76),
iridium (Ir, $Z =$~77), 
platinum (Pt, $Z =$~78),
gold (Au, $Z =$~79), 
lead (Pb, $Z =$~82), and
bismuth (Bi, $Z =$~83)
\citep{cowan96,cowan02,cowan05,sneden98,sneden03,denhartog05,
roederer09,roederer10b,roederer12a,
barbuy11,peterson11,roederer12b,roederer12c}.
Here, we employ new NUV STIS spectra with high signal-to-noise
(S/N) ratios to push abundance studies farther into the NUV
in bright, metal-poor red giant stars.

Throughout this paper we use
the standard definitions of elemental abundances and ratios.
For element X, the logarithmic abundance is defined
as the number of atoms of element X per 10$^{12}$ hydrogen atoms,
$\log\epsilon$(X)~$\equiv \log_{10}(N_{\rm X}/N_{\rm H}) +12.0$.
For elements X and Y, the logarithmic abundance ratio relative to the
solar ratio of X and Y is defined as
[X/Y]~$\equiv \log_{10} (N_{\rm X}/N_{\rm Y}) -
\log_{10} (N_{\rm X}/N_{\rm Y})_{\odot}$.
Abundances or ratios denoted with the ionization state
indicate the total elemental abundance as derived from 
that particular ionization state
after ionization corrections have been applied.
When reporting relative abundance ratios (e.g., [X/Fe]),
these ratios are constructed
by comparing the abundance of element X derived from the
neutral species with the iron abundance derived
from Fe~\textsc{i} and abundance derived from the 
ionized species of element X with the iron abundance
derived from Fe~\textsc{ii}.

\section{Observations}
\label{observations}

\subsection{Target Selection}

Stars with strong or moderate levels of \rpro\ material have been
well-studied using the GHRS and STIS.
Our goal in the present study is to investigate the abundances of
heavy elements in stars with modest levels of \rpro\ material.
We select targets that are among the brightest
metal-poor stars in the sky.
We restrict ourselves to red giants, where the line opacity of
potentially weak absorption lines will not be completely 
overwhelmed by the continuous opacity, as would be the case
for warmer subgiants or main-sequence stars.
We select stars with a range of \rpro\ enrichment levels
as indicated from previous ground-based studies.
Our final sample is comprised of four stars.
\hdb\ is the brightest metal-poor star in the sky and has 
been observed previously with STIS.
We include it in our program
to increase the S/N at shorter wavelengths.
\hdc\ has been observed previously with the GHRS, though with
only very limited wavelength coverage.
These are the first high-resolution NUV spectral observations
of \hda\ and \hdd.

\subsection{HST/STIS Spectra}

\input{tab1}

\input{tab2}

In Program GO-12268, we obtained
new STIS \citep{kimble98,woodgate98} observations using
the E230M echelle grating, centered on 2707\,\AA, and the
NUV Multianode Microchannel Array (MAMA)
detector.
The 0\farcs06\,$\times$\,0\farcs2 slit 
yields a $\sim$~2~pixel 
resolving power (R~$\equiv \lambda/\Delta\lambda$) $\sim$~30,000.
This setup produces wavelength coverage from 2280\,\AA--3115\,\AA\ in a single
exposure.
Table~\ref{stistab} presents the log of observations acquired with STIS.
The observations are taken in a standard sequence that includes
acquisition and peak-up images to center the star on the narrow slit.
These observations are then reduced and 
calibrated using the standard \textit{calstis} pipeline.

We also use existing STIS spectra of \hdb\ to supplement our 
new observations. 
These observations were taken as part of Program GO-8111,
and they have been discussed in detail in \citet{cowan05}.
These spectra were taken with the same instrument setup and
have been co-added with the new observations.

After STIS was restored during \textit{HST} Servicing Mission 4 in 2009,
the NUV MAMA detector exhibited 
a significantly elevated dark current.
This is likely caused by 
charged particles from the South-Atlantic Anomaly 
exciting meta-stable states in the detector impurities.
When these states become thermally excited, 
they decay to the ground state by 
producing ultraviolet (UV)
photons that are detected by the MAMA photocathode
\citep{kimble98,zheng11}.
This reduces the S/N 
that can be attained 
to levels lower than would otherwise be expected from 
Poisson statistics.
The S/N estimates listed in Table~\ref{sntab} reflect this reality.
The dark current is significantly lower in the
STIS spectra of \hdb\ taken from Program GO-8111,
and Poisson statistics dominate the S/N budget in these observations.
Nevertheless,
our new observations of \hdb\ increase the S/N of the
previous observations by 50\%--60\%.

\subsection{Magellan/MIKE Spectra}

We supplement our STIS observations with high-resolution optical
spectra obtained from the ground using the
Magellan Inamori Kyocera Echelle (MIKE) spectrograph \citep{bernstein03}
on the Magellan-Clay Telescope at Las Campanas Observatory.
The MIKE spectra were taken with the 0\farcs7\,$\times$\,5\farcs0 slit, 
yielding
a resolving power of $R \sim$~41,000 in the blue 
and $R \sim$~35,000 in the red, split by a dichroic
around 4950\,\AA.
This setup achieves complete wavelength coverage from 
3350\,\AA--9150\,\AA. 
Data reduction,
extraction, and wavelength calibration were performed using 
the current version of the MIKE data reduction pipeline
(written by D.\ Kelson; see also \citealt{kelson03}).
Coaddition and continuum normalization were performed within the 
Image Reduction and Analysis Facility (IRAF) 
environment, and final S/N estimates are listed in Table~\ref{sntab}.

\subsection{Keck/HIRES Spectra}

To fill in the remaining spectral gap between the STIS and MIKE 
spectra, we use data collected with the High Resolution Echelle
Spectrometer (HIRES; \citealt{vogt94}) on the Keck~I Telescope
on Mauna Kea.  
These spectra were taken with the 0\farcs86\,$\times$\,7\farcs0 slit, 
yielding a resolving power of $R \sim$~45,000.
This setup achieves complete wavelength coverage from
3120\,\AA--4640\,\AA.
Results from some of these spectra have been published previously
in \citet{cowan05}, and
we refer the interested reader to this paper for further details.
S/N estimates are given in Table~\ref{sntab}.

\section{Neutron-Capture Transitions in the NUV}
\label{transitions}

We begin our search for useful transitions of heavy elements 
using line lists published in previous investigations.
We supplement these lists with low-excitation lines provided
in the compilation by \citet{morton00}.
We also include low-excitation lines or lines with relatively 
large transition probabilities. 
We include possible transitions of both the neutral and first ion
species, though usually one ionization state of each
element dominates in these cool stellar atmospheres
(see, e.g., Figure~1 of \citealt{roederer12b}).
Any NUV transitions of elements
with several optical lines that yield reliable abundances
have been ignored for the
purposes of this study (e.g., the rare earth elements).
We retain all potentially useful
zirconium (Zr, $Z =$~40) transitions in the NUV
as a control to compare the optical and NUV abundance scales.

We also search for heavy-element transitions by comparing
the spectra of \hda\ and \hdd.
These two stars have nearly identical effective temperatures and
metallicities.
Differing abundances of heavy elements in their atmospheres
account for most of the differences in their spectra.
We consider extra absorption detected in \hda\ relative to \hdd\ 
possible evidence of a heavy-element transition, and we attempt to
match these wavelengths with lines in the 
National Institute of Standards and Technology (NIST) 
Atomic Spectra Database \citep{ralchenko11}
or other laboratory studies.
This technique 
identified several lines of interstellar absorption
towards \hdd\ (see Section~\ref{reddening}) and
yielded a few heavy-element transitions worth 
pursuing further.

\begin{figure}
\includegraphics[angle=0,width=3.4in]{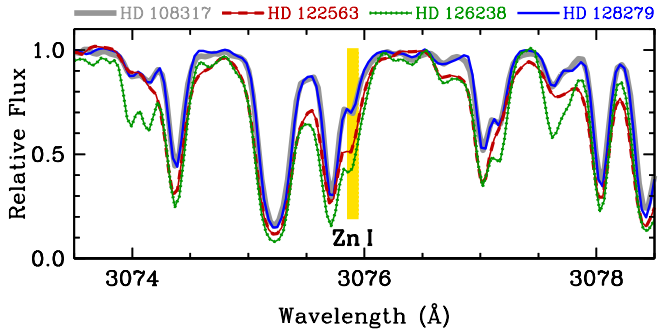}
\caption{
\label{znplot}
STIS spectra of all four stars around the Zn~\textsc{I} 3075\,\AA\ line,
marked by the shaded region.
 }
\end{figure}

\begin{figure}
\includegraphics[angle=0,width=3.4in]{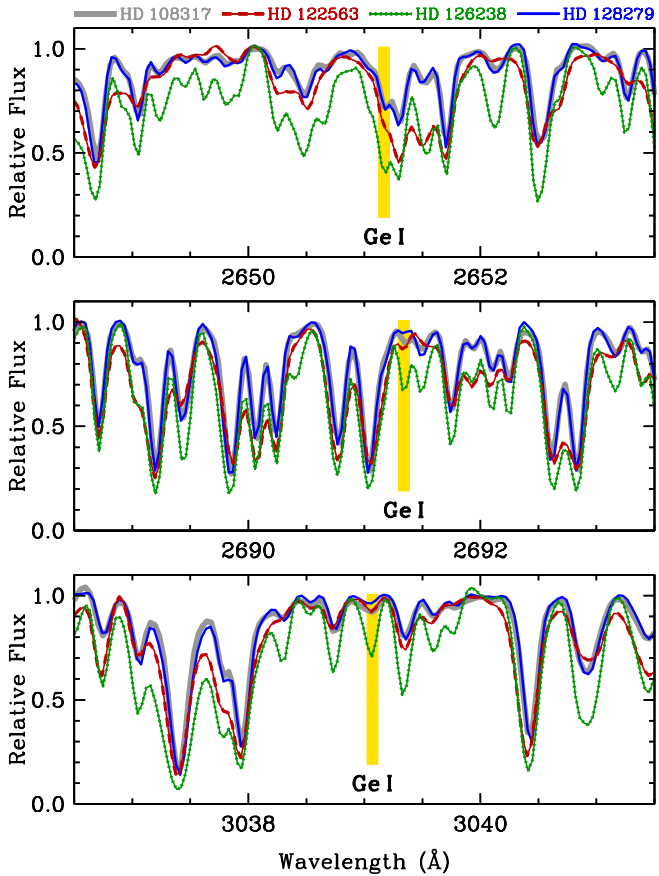}
\caption{
\label{geplot}
STIS spectra of all four program stars around the 
Ge~\textsc{i} 2651\,\AA, 2691\,\AA, and
3039\,\AA\ lines.
 }
\end{figure}

\begin{figure}
\includegraphics[angle=0,width=3.4in]{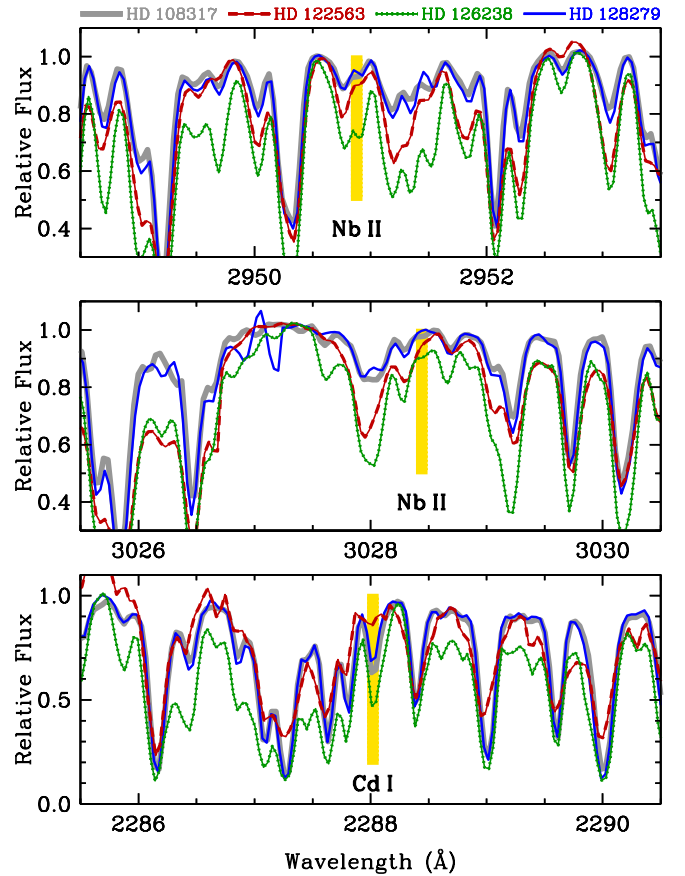}
\caption{
\label{nbplot}
STIS spectra of all four program stars around the 
Nb~\textsc{ii} 2950\,\AA\ and 3028\,\AA\
lines and the Cd~\textsc{i} 2288\,\AA\ line.
 }
\end{figure}

\begin{figure}
\includegraphics[angle=0,width=3.4in]{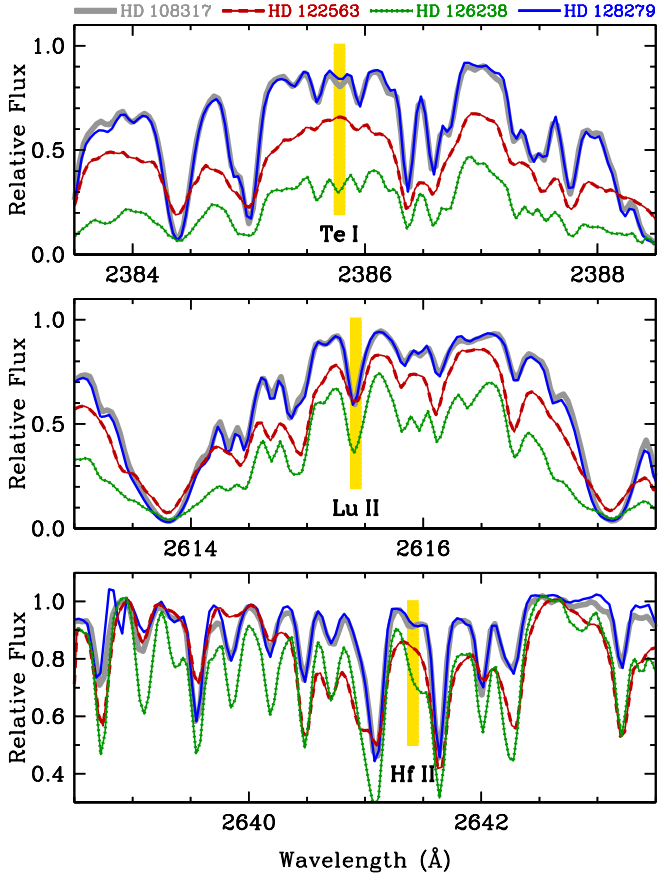}
\caption{
\label{teplot}
STIS spectra of all four program stars 
around the Te~\textsc{i} 2385\,\AA\ line,
the Lu~\textsc{ii} 2615\,\AA\ line,
and the Hf~\textsc{ii} 2641\,\AA\ line.
 }
\end{figure}

\begin{figure}
\includegraphics[angle=0,width=3.4in]{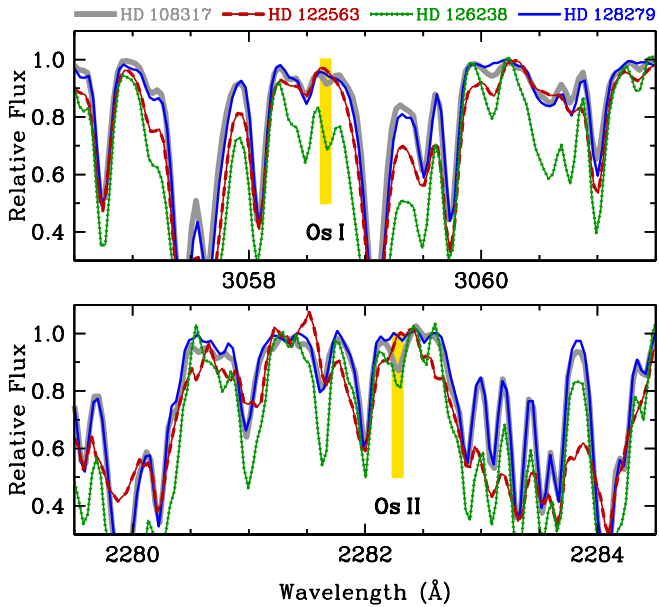}
\caption{
\label{osplot}
STIS spectra of all four program stars 
around the Os~\textsc{i} 3058\,\AA\ line
and the Os~\textsc{ii} 2282\,\AA\ line.
 }
\end{figure}

\begin{figure}
\includegraphics[angle=0,width=3.4in]{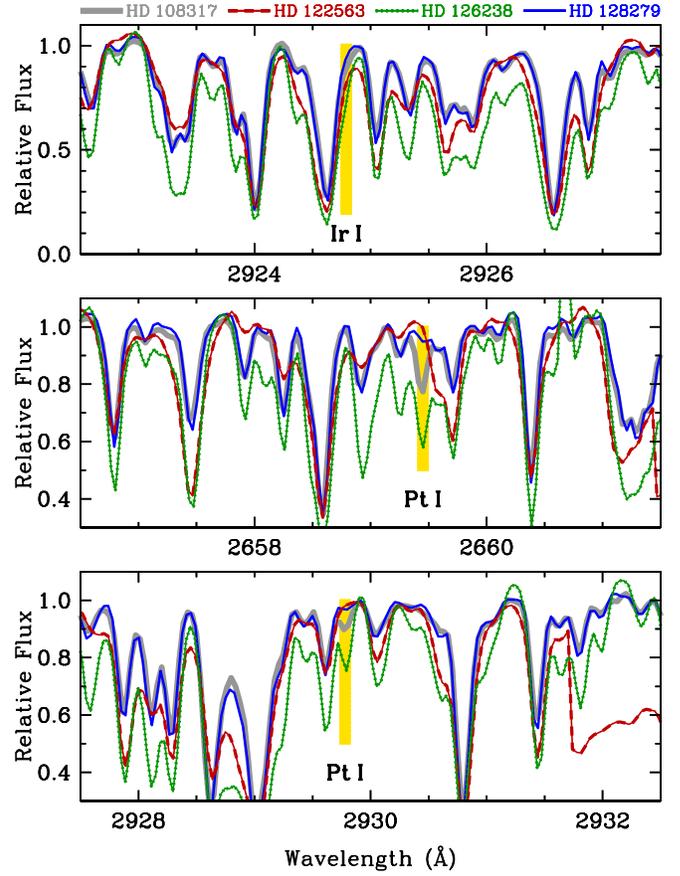}
\caption{
\label{irplot}
STIS spectra of all four program stars 
around the Ir~\textsc{i} 2924\,\AA\ line
and the Pt~\textsc{i} 2659\,\AA\ and 2929\,\AA\ lines.
 }
\end{figure}

\begin{figure}
\includegraphics[angle=0,width=3.4in]{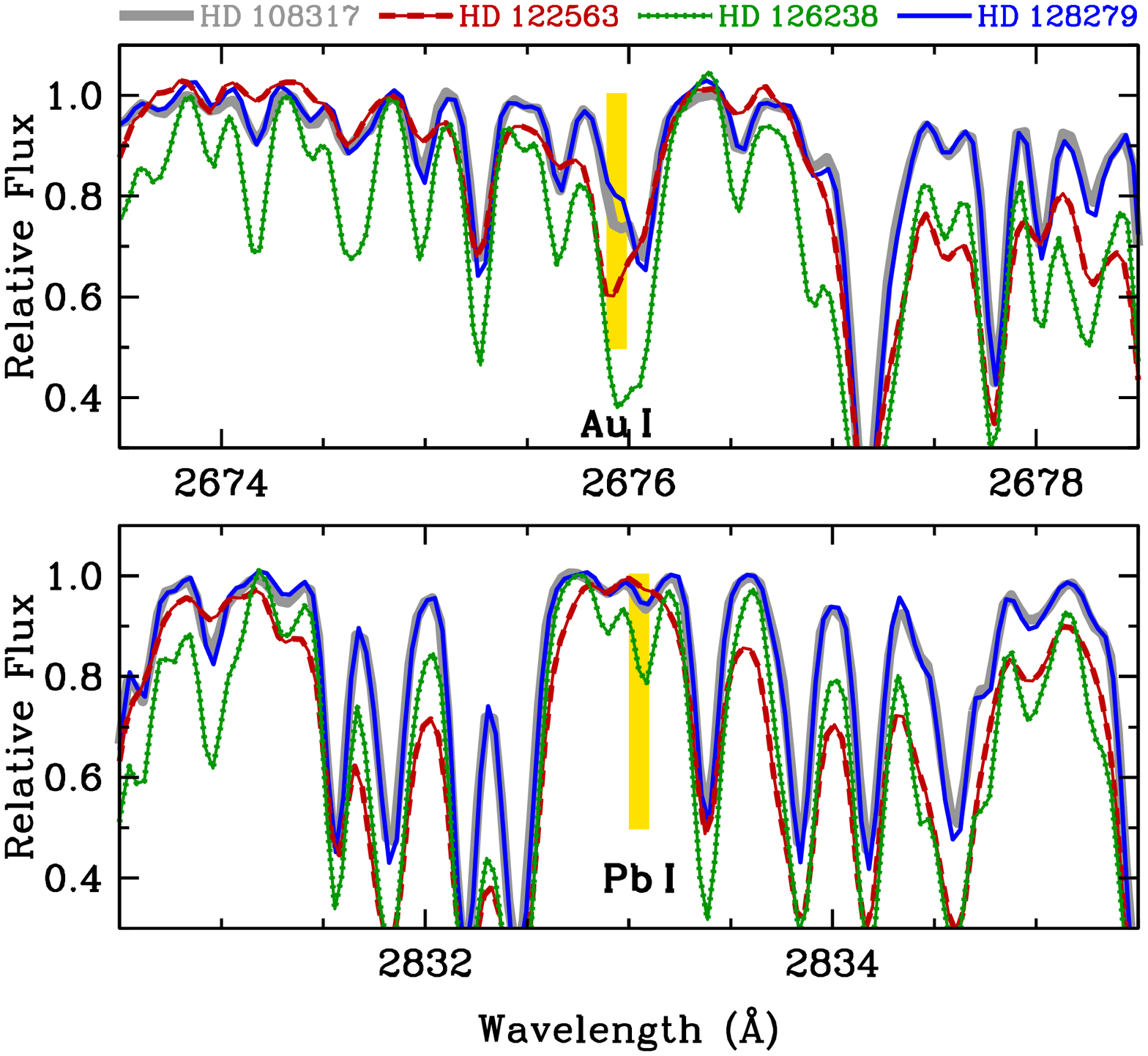}
\caption{
\label{auplot}
STIS spectra of all four program stars 
around the Au~\textsc{i} 2675\,\AA\ line
and the Pb~\textsc{i} 2833\,\AA\ line.
 }
\end{figure}

Our search produced an initial set of approximately 300~transitions
between 2280\,\AA\ and 3115\,\AA\ for further consideration.
We refine this list by examining the spectra of \hda\ and \hdd\ 
at the wavelength of each transition.
In a few cases no absorption is detected at all.
Many potentially useful heavy-element transitions lie in regions 
that are hopelessly 
blended with absorption from more abundant species,
usually neutral or first ion iron group elements.
We discard these transitions from our list.
This step reduces our initial list by about half.

We next generate a synthetic spectrum covering $\pm$~3\,\AA\ around
each line of interest to determine whether the
absorption is due to the species of interest.
We produce line lists from the \citet{kurucz95} atomic and
molecular lists, updating them with laboratory $\log(gf)$ values 
when known.
We include hyperfine splitting (hfs) structure
for odd-$Z$ iron group elements.
This step
reveals that many of the potential heavy-element transitions are
far too weak or too blended to be reliably detected in our sample.

We are left with 27~lines worthy of close scrutiny.
These lines are discussed in detail in Appendix~\ref{appendix}.
In addition to a handful of Zr~\textsc{ii} lines, 
our STIS spectra
reveal absorption lines due to the heavy elements
zinc (Zn, $Z =$~30),
germanium, 
niobium (Nb, $Z =$~41),
cadmium, 
tellurium, 
lutetium, 
hafnium (Hf, $Z =$~72),
osmium, 
iridium, 
platinum, 
gold, and 
lead. 
The spectral regions surrounding lines of each of these elements
are shown for all four stars in 
Figures~\ref{znplot} through \ref{auplot}.
In addition, we derive upper limits from non-detections of 
molybdenum (Mo~\textsc{ii}; we can detect Mo~\textsc{i} in
the ground-based spectra) and
bismuth.

\input{tab3-fauxstub}

In Table~\ref{atomictab} 
we present a list of 229~lines examined in each star, 
including the 27~lines examined in our new STIS spectra.
This table includes the line wavelength, species identification,
excitation potential (E.P.) of the lower level of the transition, 
$\log(gf)$ value, and references to the
source of the $\log(gf)$ value and any hfs
structure or isotope shifts (IS) included in the syntheses.

The rare earth elements from lanthanum (La, $Z =$~57)
through lutetium 
are exclusively detected as first ions in the atmospheres
of metal-poor red giant stars.
The first ionization potentials of these elements are relatively
low, ranging from 5.43~eV to 6.25~eV.
In contrast, the elements at the third \rpro\ peak (osmium
through gold) and lead
have significantly higher first ionization potentials, ranging from
7.42~eV to 9.23~eV.
The neutral species of these elements are usually detected,
though the ionized species may also be present in substantial amounts.
Strontium through cadmium and 
tellurium have gradually increasing first ionization
potentials from 5.69~eV to 9.01~eV.
Strontium, yttrium, zirconium, and niobium are detected as
first ions in these stars, while
molybdenum through cadmium and tellurium are detected in their neutral states.
Germanium has a high first ionization potential, 7.90~eV, and is
also detected in the neutral state.

\section{Iron Equivalent Widths}
\label{ew}

\input{tab4-stub}

We measure equivalent widths (EWs) of Fe~\textsc{i} and
\textsc{ii} lines from our MIKE spectra
using a semi-automatic routine that 
fits Voigt absorption line profiles
to continuum-normalized spectra.
The complete list of EW measurements is given in Table~\ref{fetab},
which is available only in the online edition of the journal.
A sample is shown in the printed edition to demonstrate
its form and content.

\section{Model Atmospheres}
\label{modelatm}

\input{tab5}

Our target stars are all relatively nearby at distances of 150~pc to
300~pc.
The Hipparcos Satellite \citep{perryman97} measured parallaxes, 
$\pi$,
to each of these stars to better than 20\%,
as given by the reduction validated by \citet{vanleeuwen07}.
Table~\ref{datatab} presents the Hipparcos parallax, 
distance, 
$V$~magnitude (from SIMBAD), 
$K$~magnitude (from the Two Micron All Sky Survey, \citealt{skrutskie06}), 
reddening (from the \citealt{schlegel98} dust maps, 
modified in cases of high reddening according to the prescription
given in \citealt{bonifacio00}), and 
de-reddened $V-K$ color (assuming the extinction coefficients of 
\citealt{cardelli89}). 
We use the $V-K$ color-temperature relation derived by \citet{alonso99b}
to compute an initial estimate of effective temperature (\teff), 
where the quoted statistical uncertainties account for uncertainties in
the photometry and the scatter in the color-\teff\ relation.
We compute initial estimates of the surface gravities (\logg) 
using these data, bolometric corrections from \citet{alonso99a},
an assumed stellar mass of 0.8~$M_{\odot}$, and
solar parameters $M_{\rm bol} =$~4.74,
\logg$_{\odot} =$~4.44, and
\teff$_{\odot} =$~5780~K.
The quoted statistical uncertainties on \logg\ in
Table~\ref{datatab} reflect the uncertainties
of the input quantities.

We derive abundances of iron from Fe~\textsc{i} and \textsc{ii} using our 
measured EWs, interpolations among 
the $\alpha$-enhanced grid of ATLAS9 model atmospheres
\citep{castelli04}, and
the latest version of the analysis code MOOG \citep{sneden73}.
This version of MOOG
includes the contribution of Rayleigh scattering from atomic H~\textsc{i}
in the source function \citep{sobeck11}.

\input{tab6}

For reasonable estimates of the microturbulent velocity (\vt),
each initial temperature estimate produces a strong correlation
of iron abundance (derived from Fe~\textsc{i}) with E.P.,
implying that the photometric temperatures are too warm by 
several hundred K.
We adjust the temperature and microturbulent velocity to minimize abundance 
correlations 
between E.P.\ and line strength, $\log$(EW/$\lambda$),
respectively.
We set the overall model metallicity to the 
iron abundance derived from Fe~\textsc{ii}.
We then recompute \logg\ as described above and iterate until
all four parameters converge. 
Our adopted model parameters are listed in Table~\ref{modeltab}.
This method does not enforce iron ionization balance,
so the iron abundance derived from Fe~\textsc{i} lines 
is not necessarily equal to 
the iron abundance derived from Fe~\textsc{ii} lines.

A recent measurement of the radius of \hdb\ results in a 
derived \teff\ (4598~$\pm$~41~K; \citealt{creevey12})
that is intermediate between
the photometric (4680~$\pm$~65~K) and
spectroscopic (4450~K)
values.
The corresponding value of \logg\ (1.60~$\pm$~0.04; \citeauthor{creevey12})\
agrees with the value derived from the Hipparcos parallax
(1.58~$\pm$~0.13), both of which are slightly higher
than our derived value (1.37).
This comparison and a comparison of the values in
Tables~\ref{datatab} and \ref{modeltab} suggest that
systematic
uncertainties in \teff\ may be $\approx$~4\%--6\% and
systematic
uncertainties in \logg\ may be $\approx$~20\%.
This offset between photometrically and
spectroscopically determined model parameters is well-known
(e.g., \citealt{frebel10b}).
Our work simply reaffirms this offset.

\section{Interstellar Absorption}
\label{reddening}

In principle,
overestimation of the reddening can account for 
the large differences between the photometric and spectroscopic temperatures.
The \citet{schlegel98} dust maps predict significant
color excess at infinity, $E(B-V) > 0.1$,
along the lines of sight to \hdc\ and \hdd.
These two stars are located at relatively low Galactic latitude
($b < 30^{\circ}$) and may be within the reddening layer.
They also have large temperature adjustments ($-$320~K and $-$470~K,
respectively).
We detect probable interstellar absorption from transitions from
the ground states of
Na~\textsc{i} (5889.95\,\AA, 5895.92\,\AA), 
K~\textsc{i} (7698.96\,\AA),
Mn~\textsc{ii} (2593.73\,\AA, 2605.69\,\AA), and 
Fe~\textsc{ii} (2343.50\,\AA, 2373.74\,\AA, 2382.04\,\AA,
2585.88\,\AA, 2599.40\,\AA)
in the spectrum of \hdd.
These interstellar lines all have a velocity 
offset of about $+$60~\kmsec\ relative to the stellar lines.
We also detect interstellar absorption in the spectrum of \hdc\ 
due to Na~\textsc{i}.
The lines of sight to 
\hda\ and \hdb\ are less reddened,
$E(B-V) < 0.03$, according to the maps of \citet{schlegel98}.
These stars are located at higher Galactic latitude ($b > 65^{\circ}$)
and have lower temperature corrections ($-$160~K and $-$230~K, 
respectively).

We use the interstellar Na~\textsc{i} 
lines to estimate the reddening towards these stars.
We make an approximate removal of the telluric water vapor lines
in this region using a smoothed version of the
the telluric spectrum 
presented by \citet{hinkle00}.
We do not detect Na~\textsc{i} interstellar absorption towards \hda\ and \hdb.
We can only estimate the interstellar absorption from the
Na~\textsc{i} 5895\,\AA\ line towards \hdc\ due to
residual telluric contamination of the 5889\,\AA\ line.
Both lines are cleanly detected towards \hdd, but they are not
resolved in our spectra.
Their
EWs differ by a factor of two within the uncertainties
(70~$\pm$~1~m\AA\ and 34~$\pm$~1~m\AA),
which matches the ratio of their $f$-values and suggests that
neither line is saturated.
We calculate column densities
$\log N$~(Na~\textsc{i})~$=$~11.76~$\pm$~0.01~cm$^{-2}$ toward \hdc\ and
$\log N$~(Na~\textsc{i})~$=$~11.55~$\pm$~0.01~cm$^{-2}$ toward \hdd.
Using the linear relationship between $\log N$~(Na~\textsc{i}) and 
$\log N$~(H~\textsc{i}~$+$~H$_{2}$)~$= \log N$~(H)
found by \citet{ferlet85}, we assume a common Na/H ratio and Na
depletion in
the solar neighborhood to estimate
$\log N$~(H)~$=$~20.05~cm~$^{-2}$ towards \hdc\ and
$\log N$~(H)~$=$~19.85~cm~$^{-2}$ towards \hdd.
Upper limits from the non-detection of interstellar Na~\textsc{i}
towards \hda\ and \hdb\ imply 
$\log N$~(Na~\textsc{i})~$<$~10.71~cm$^{-2}$ and
$\log N$~(H)~$<$~19.04~cm~$^{-2}$.

Using the mean relationship between 
$N$~(H~\textsc{i}~$+$~H$_{2}$) and $E(B-V)$ derived by \citet{bohlin78}, 
we would infer $E(B-V)$ between 0.01 and 0.02 towards \hdc\ and \hdd\
and even less towards \hda\ and \hdb.
This is substantially less than the amount of reddening predicted 
by the \citet{schlegel98} dust maps.
We would only expect
$\approx$~40\,K--60\,K of a decrease in \teff\ based on these values,
which is far short of the adjustments found necessary to produce
no correlation between iron abundance and E.P.
We note, however, that there is more than a factor of 5
in the scatter
between $\log N$~(Na~\textsc{i}) and $\log N$~(H)
at these column densities,
and the relationship between $N$~(H~\textsc{i}~$+$~H$_{2}$) and $E(B-V)$
is poorly defined at low reddening.
The actual amount of gas along the lines of sight to \hdc\ and \hdd\
could be higher by a factor of a few, which would bring the
\teff\ adjustments into line with those found for the 
unreddened stars \hda\ and \hdb.

\begin{figure}
\includegraphics[angle=0,width=3.4in]{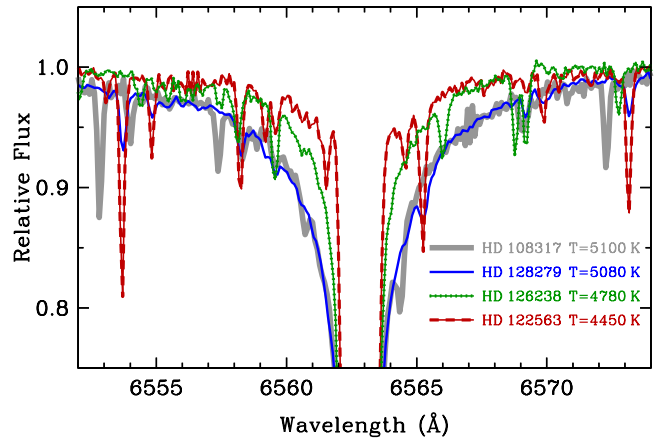}
\caption{
\label{halphaplot}
Comparison of the H$\alpha$ line profiles for the four program stars.
For reference we list the
temperatures derived by minimizing the Fe~\textsc{i} abundance
correlation with E.P.
The two warmest stars, \hda\ and \hdd,
have nearly identical H$\alpha$ profiles.
\hdc\ and \hdb\ are cooler, and
the H$\alpha$ line profiles confirm the spectroscopic
temperatures in a relative sense.
}
\end{figure}

Other
comparisons of the spectra suggest that this explanation is plausible. 
Figure~\ref{halphaplot} illustrates the profiles of the
H$\alpha$ line in all four stars.
The line is broadest and virtually indistinguishable
in \hda\ and \hdd.
\hdc\ and \hdb\ have narrower profiles.
Our derived model parameters for \hda\ and \hdd\ are 
identical within the uncertainties, and
Figures~\ref{znplot} through \ref{auplot} 
demonstrate how similar the spectra of these two stars are.
This confirms the relative sense of our spectroscopic temperatures, and it
supports our decision to adjust the temperatures from the
initial photometric estimates to a spectroscopic scale.

\section{Abundance Analysis}
\label{abundances}

\input{tab7}

\input{tab8}

We perform the abundance analysis by comparing the observed
spectrum with synthetic spectra calculated using MOOG.
Sources for the atomic data for lines of interest are presented
in Table~\ref{atomictab}.
We use damping constants from \citet{barklem00}
and \citet{barklem05b}, when available, and otherwise
we resort to the standard \citet{unsold55} approximation.
We adopt the S.S.\ isotopic fractions for Cu~\textsc{i} and 
the \rpro\ isotopic fractions reported in \citet{sneden08}
for all other elements 
where multiple isotopes are considered in the calculation
(Ba~\textsc{ii},
Nd~\textsc{ii},
Sm~\textsc{ii},
Eu~\textsc{ii},
Yb~\textsc{ii},
Ir~\textsc{i},
Pt~\textsc{i}, and 
Pb~\textsc{i}).

When our lines of interest are mildly blended with other absorption
features for which no laboratory $\log(gf)$ value is available, 
we treat the strength of the blending feature as a free parameter.
When the strength of blending features is completely degenerate with the
strength (abundance) of lines of interest, we discard
these lines from consideration.
Some unidentified absorption features are detected in our stellar spectra
but not present in the line lists, 
and many lines in the UV lack reliable transition probabilities.
This is a well-known problem in the UV (e.g., \citealt{leckrone99}).
When no transition in our line list can
be reasonably adjusted to account for this absorption, we
assume the absorption is due to an uncatalogued Fe~\textsc{i} line
with a lower excitation level of 1.5~eV
and treat the $\log(gf)$ value as a free parameter in our synthesis.
This technique has been successfully applied 
previously by, e.g., \citet{peterson11}.

Our HIRES and MIKE spectra overlap in wavelength coverage.
We find no significant difference between abundances derived from
the HIRES and MIKE spectra for these stars.
When a line is present in both, we adopt the abundance derived
from the higher resolution HIRES spectrum.

Tables~\ref{finalabund1} and \ref{finalabund2} present the
absolute and relative abundances, statistical and
total uncertainties, and the number of 
lines examined in each star.
The reference solar abundances listed in these tables
are taken from \citet{asplund09}.
Mean abundances are weighted by the statistical 
uncertainties of each line. 
The statistical uncertainties quoted in these tables
include components from the synthetic spectra fitting,
uncertainties in the individual \loggf\ values, 
and uncertainties from the wavelength-dependent 
corrections discussed in Section~\ref{corrections}.
We assume the minimum uncertainty per line is equivalent to
the standard deviation of the well-measured species Zr~\textsc{ii}
for each star (0.11--0.17~dex).
Lines whose fit is poorer due to significant blending or
uncertainty in the continuum identification are marked
with a colon in Table~\ref{atomictab} and given lower weight in the average.
Systematic uncertainties for each of neutral and singly-ionized species
have been discussed in more detail by \citet{cowan05}, and we 
adopt their assessment of the uncertainties.
These uncertainties amount to 0.12~dex and 0.17~dex for 
neutral and ionized species, respectively, and are added
in quadrature with the statistical uncertainties to 
form the total uncertainties listed in Tables~\ref{finalabund1} and 
\ref{finalabund2}.

The dominant sources of continuous opacity 
in these stars at the wavelengths considered
are H$^{-}$ bound-free absorption and Rayleigh
scattering from neutral H.
In metal-rich stars, metal ionization makes significant
contributions to the continuous opacity, but this is insignificant in
the metal-poor stars considered here
(cf.\ \citealt{roederer12c}).

\section{Abundance Trends with Wavelength}
\label{corrections}

\begin{figure}
\includegraphics[angle=0,width=3.4in]{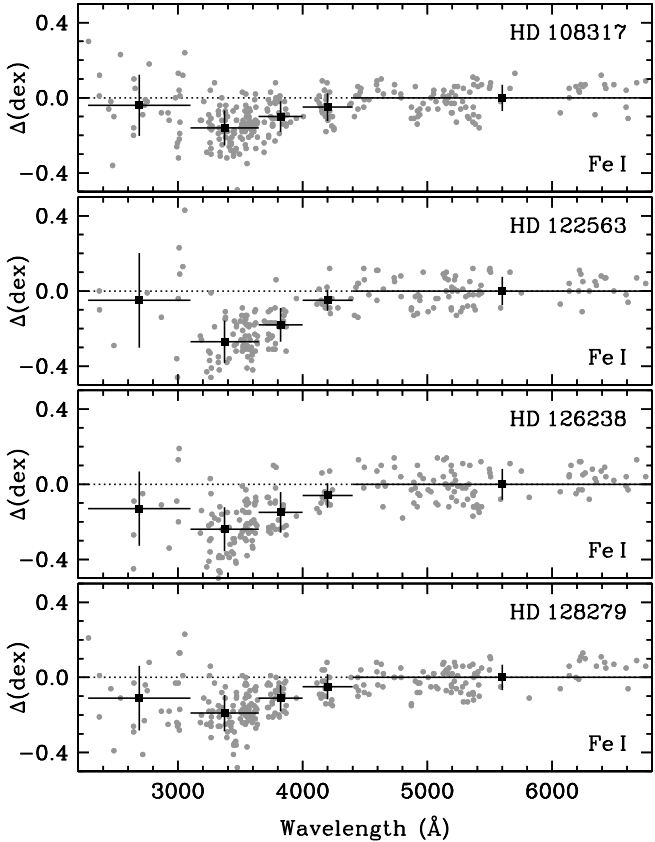}
\caption{
\label{ironplot}
Derived iron abundances from Fe~\textsc{i} lines as a function of wavelength.
The gray dots represent abundances derived from individual lines.
Large black squares represent the mean abundance in each 
wavelength interval.
Vertical error bars indicate the standard deviation, and
horizontal error bars mark the wavelength interval.
The zeropoint for each star is defined as the 
iron abundance derived from Fe~\textsc{i} lines longward of 4400\,\AA,
which is marked by the dotted line in each panel.
 }
\end{figure}

We measure EWs of 119--134 Fe~\textsc{i} and \textsc{ii} lines,
spanning 3765\,\AA---6750\,\AA,
in each star in our sample (see Table~\ref{fetab}).
At the referee's suggestion, we also examine Fe~\textsc{i}
and \textsc{ii} lines at wavelengths as short as 2283\,\AA.
We derive the abundances of all iron lines not listed in Table~\ref{fetab}
by spectrum synthesis since these additional lines lie in 
more crowded spectral regions.
The final list of iron abundances is presented
in Table~\ref{irontab},
which is available only in the online edition of the journal.
A sample is shown in the printed edition to demonstrate
its form and content.

Figure~\ref{ironplot} illustrates the iron abundances
derived from Fe~\textsc{i},
by far the species with the most lines considered
(211--300~lines per star).
Significant variations in Fe~\textsc{i} 
persist at short wavelengths.
These variations are characterized as a decrease in the 
average abundance derived from Fe~\textsc{i} lines at short wavelengths
compared to lines at long wavelengths.
When compared with abundances derived from Fe~\textsc{i} lines
at wavelengths longward of 4400\,\AA, Fe~\textsc{i} lines
short of $\approx$~3100\,\AA\ and between 4000\,\AA\ and
4400\,\AA\ yield average abundances lower by small amounts, 
typically 0.04~dex to 0.06~dex,
but as large as 0.13~dex.
Iron abundances derived from Fe~\textsc{i} lines with wavelengths between
$\approx$~3100\,\AA\ and $\approx$~4000\,\AA\ show an even larger
deficiency of 0.10~dex to 0.27~dex.

\begin{figure}
\includegraphics[angle=270,width=3.4in]{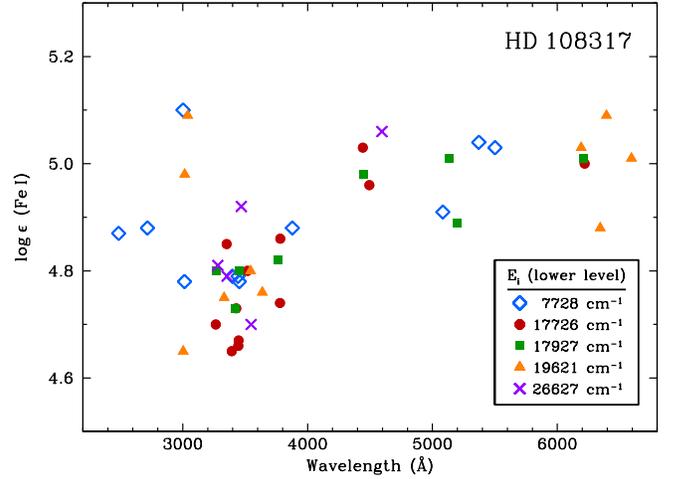}
\caption{
\label{levelplot}
Derived iron abundances from Fe~\textsc{i} lines 
in \hda\ as a function of wavelength
for select lower levels.
 }
\end{figure}

\begin{figure*}
\begin{center}
\includegraphics[angle=270,width=4.8in]{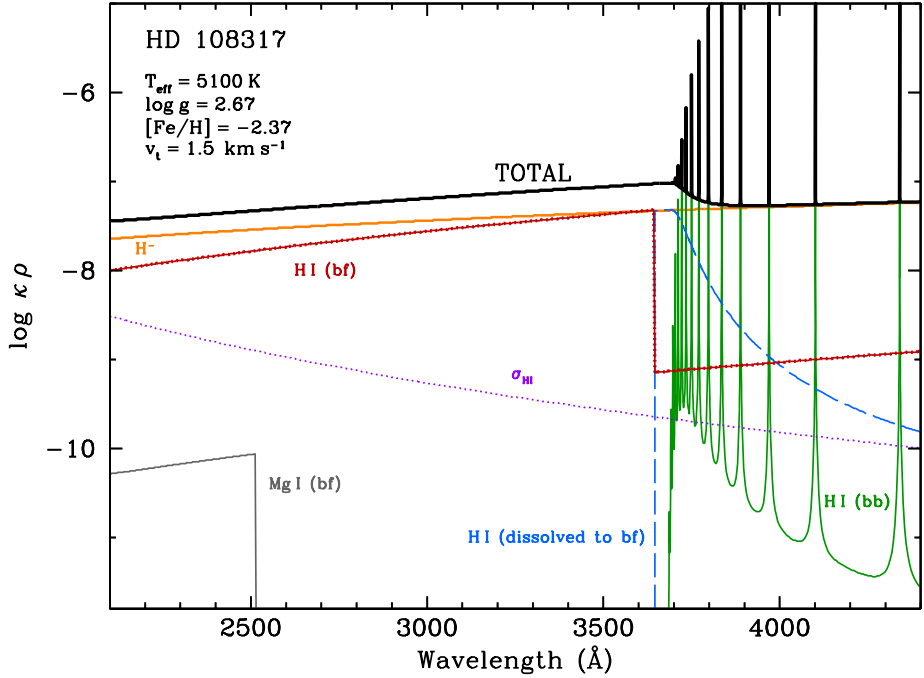}
\end{center}
\caption{
\label{opacplot}
Contributions to the continuous opacity in \hda.
The ordinate shows (logarithmic) opacity times density in units of cm$^{-1}$
at a layer in the atmosphere near 
$\tau_{\rm 5000} \sim$~1.
The bold black line indicates the total opacity contributions
from all sources.
The studded red line indicates the bound-free opacity from H~\textsc{i}.
The dashed blue line indicates the opacity from dissolved 
states of H~\textsc{i}.
The solid green line indicates the H~\textsc{i} bound-bound line opacity.
The solid orange line indicates the opacity from the H$^{-}$ ion.
The dotted purple line indicates the opacity from Rayleigh scattering
from H~\textsc{i}.
The solid gray line indicates the bound-free opacity from Mg~\textsc{i}.
 }
\end{figure*}

\begin{figure}
\includegraphics[angle=0,width=3.4in]{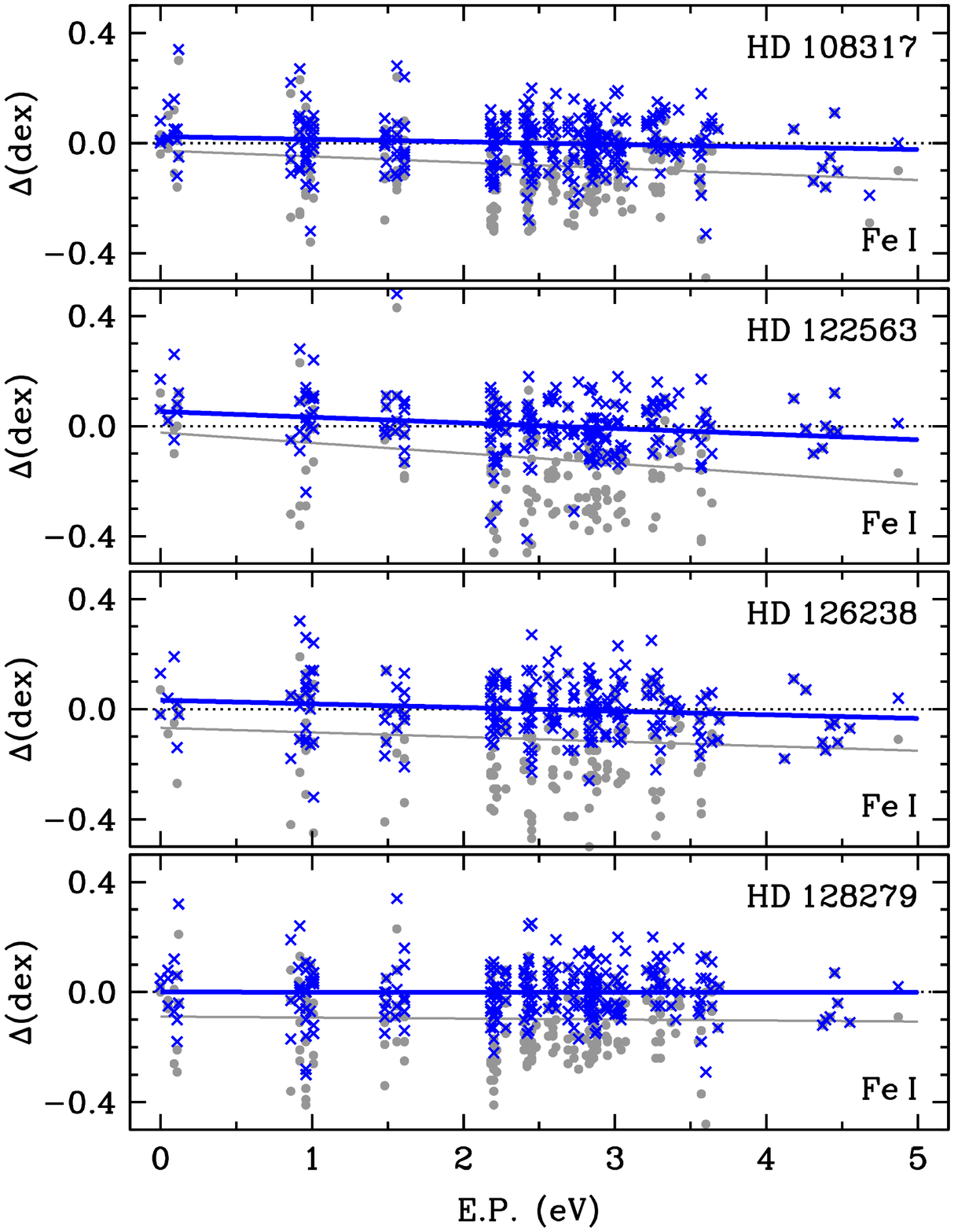}
\caption{
\label{epplot}
Line-by-line Fe~\textsc{i} abundance deviations from the corrected mean 
as a function of E.P.
The gray circles denote the uncorrected abundances, and the
blue crosses denote the corrected ones.
The gray line marks the linear least-squares fit to the 
uncorrected abundances, and the bold blue line
marks the fit to the corrected ones.
The dotted black line marks an offset of zero
from the mean iron abundance derived from Fe~\textsc{i} lines.
 }
\end{figure}

\input{tab9-stub}

All four stars qualitatively show the same effect,
although the magnitude of the ``dip'' appears to be 
larger in the cooler stars.
Our tests with warmer metal-poor subgiant stars 
suggest that they, too, exhibit a qualitatively similar feature
at these wavelengths.
Omitting lines with the largest transition probability
uncertainties does not change the result appreciably.

If departures from LTE in neutral iron are the source of this effect,
we might expect lines originating from different lower levels
to behave differently.
Figure~\ref{levelplot} demonstrates that this is not the case for \hda,
the warmest star in our sample, and
the results are similar for \hdb, the coolest star in our sample.
Figure~\ref{levelplot} 
shows the iron abundance derived from Fe~\textsc{i} lines
as a function of wavelength for five selected lower levels.
These levels are chosen to have a reasonably large set of 
high-quality \loggf\ values both within and outside the
affected wavelength region.
The levels range from 0.96~eV to 3.3~eV, and all show
the same effect of producing low iron abundances between
3100\,\AA\ and 4000\,\AA.
This indicates that we are not observing a non-LTE distribution 
of level populations masquerading as a wavelength-dependent trend.

An underestimate of the continuous opacity in the region of the dip
would lead to spuriously low abundance results relative to 
shorter and longer wavelengths.
This would also affect abundances derived from
lines of all species, not just Fe~\textsc{i},
and we see hints of this in the abundances derived
from, e.g., Fe~\textsc{ii} or Zr~\textsc{ii} lines.

The wavelengths of the dip hint that such an effect
could be related to the transition from the Paschen continuum 
to the Balmer continuum.
In an effort to test this as a possibility for wavelengths longer than 
the Balmer jump at 3647\,\AA,
we have modified MOOG to compute the H~\textsc{i} bound-free
continuous absorption according to the
occupation probability formalism articulated by
\citet{dappen87}, \citet{hummer88}, and \citet{hubeny94}.
This approximates the probability that a particular level is 
so strongly perturbed by interactions with other particles
that an electron excited to this level is effectively unbound,
or dissolved.
This produces a continuous, density-dependent 
distribution from bound-bound transitions to bound-free
in a pseudo-continuum for progressively higher levels
in a given H~\textsc{i} spectral series.
This ensures that ionization occurs gradually
rather than in sudden, discrete jumps at the 
series ionization edges.
Our implementation of these new calculations in MOOG 
follows that in Version~10.1 of \textit{turbospectrum} \citep{alvarez98}.

The results of these calculations for \hda\ are shown in 
Figure~\ref{opacplot}.
The (blue) dashed line and (green) solid line represent the
H~\textsc{i} dissolved opacity and bound-bound opacity, respectively.
Computation of 
the H~\textsc{i} bound-free opacity, shown by the (red)
studded line, has not been altered from the 
standard MOOG use of the ATLAS polynomial approximations
\citep{kurucz70} to the Coulomb cross sections of
\citet{karzas61}.
For this particular model atmosphere, the opacity contribution 
from the bound-free dissolved levels smoothly transitions into the
bound-free absorption short of the $n =$~2 series limit.
In cooler models, however, the calculated bound-free opacity from dissolved
levels is greater than the bound-free opacity short of the series limit,
and the reverse is true for warmer models.
In both cases, the discrepancy at the series limit is 
up to a factor of a few.  
While the increased opacity from the dissolved bound-free states
does reduce the effect of the iron abundance dip, 
this does not completely eliminate the effect at 
wavelengths longward of the $n =$~2 series limit.
Furthermore, it cannot affect lines between 3100\,\AA\ and 3647\,\AA\
and only minimally affects lines longward of $\approx$~3900\,\AA,
where a small dip is still present.

\input{tab10}

We are unable to identify the source of the discrepancy
among iron abundances derived from Fe~\textsc{i} lines 
at blue wavelengths in our spectra.
We have chosen to adopt an empirical, wavelength-dependent 
correction to the iron abundances, as shown by the black squares
in Figure~\ref{ironplot}.
These corrections are computed without the use of the new
dissolved state calculations in MOOG.
The continuous opacity at longer optical wavelengths, $\lambda >$~4400\,\AA,
is dominated by H$^{-}$ bound-free absorption,
so we adopt the iron abundance derived from Fe~\textsc{i} lines
at these wavelengths as the standard to which all other
values are corrected.
These corrections and their uncertainties are listed in 
Table~\ref{corrtab}.
We apply these corrections to abundances derived from 
all lines of other species,
and the corrections are reflected in 
the values presented in Table~\ref{atomictab}.
The corrected iron abundances derived from Fe~\textsc{ii} lines
show more scatter than the corrected iron abundances 
derived from Fe~\textsc{i} lines, and this
scatter dominates over any residual wavelength trends.

Our model atmosphere parameters 
do not change after making these corrections.
We derived \teff\ by removing
the correlation between iron abundance derived from 
Fe~\textsc{i} lines and E.P.~
Only the Fe~\textsc{i} lines with measured EWs were used in this procedure.
This set of lines lies mostly in the redder region of the spectrum
unaffected by the dip.
(Most studies that adopt a spectroscopic \teff\ scale use redder
sets of lines, also.)
Figure~\ref{epplot} illustrates 
the relationship between iron abundance derived from \textit{all}
Fe~\textsc{i} lines and E.P.~
The gray points illustrate the uncorrected abundances,
and the gray lines mark the linear least-squares fits to them.
The blue crosses illustrate the corrected abundances,
and the bold blue lines mark the fits to them.
Had we initially used all Fe~\textsc{i} lines
to derive \teff, and not just the
redder ones with measured EWs,
these data would have suggested even cooler spectroscopic temperatures.
The corrected Fe~\textsc{i} abundances yield
slopes consistent with zero (formally, the slopes are less than 2.6 times
their errors),
which supports our original spectroscopic \teff\ scale.
In summary,
this correction cannot be the explanation for the discrepancy
between the spectroscopic and photometric values of \teff\
for these stars.

These corrections account for the small differences
found between Tables~\ref{finalabund1} and \ref{finalabund2}
and the preliminary abundances presented in recent
papers by our group.
We proceed with caution and
defer further investigation of this unexpected
yet interesting effect to future work.

\section{Results}
\label{results}

\begin{figure}
\begin{center}
\includegraphics[angle=0,width=3.0in]{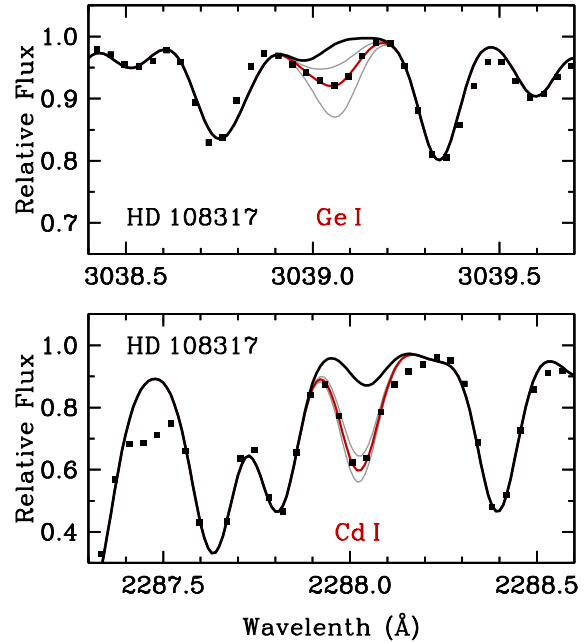}
\end{center}
\caption{
\label{synthplot1}
Comparison of observed and synthetic spectra in \hda.
The 
Ge~\textsc{i} 3039\,\AA\ and
Cd~\textsc{i} 2288\,\AA\ lines are shown.
The bold red line represents the best-fit abundance,
the thin gray lines represent variations in this abundance by $\pm$~0.30~dex,
and the bold black line represents a synthesis with no
germanium or cadmium.
}
\end{figure}

\begin{figure}
\begin{center}
\includegraphics[angle=0,width=3.0in]{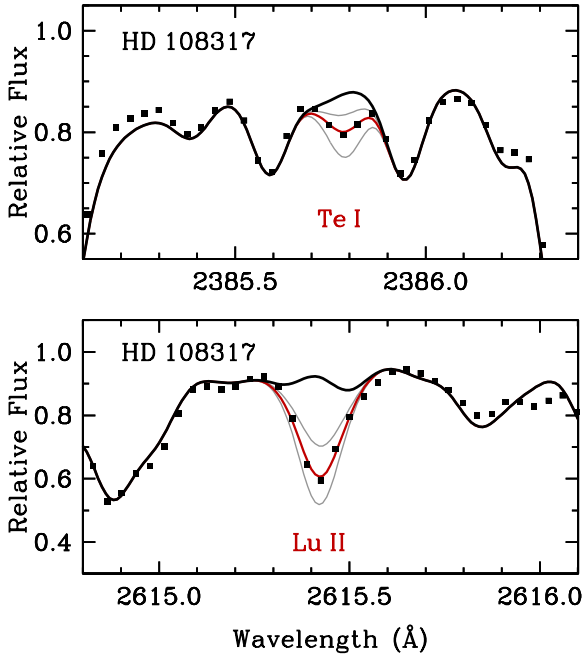}
\end{center}
\caption{
\label{synthplot2}
Comparison of observed and synthetic spectra in \hda.
The 
Te~\textsc{i} 2385\,\AA\ and
Lu~\textsc{ii} 2615\,\AA\ lines are shown.
The bold red line represents the best-fit abundance,
the thin gray lines represent variations in this abundance by $\pm$~0.30~dex,
and the bold black line represents a synthesis with no
tellurium or lutetium.
As noted in Appendix~\ref{appendix}, strong Fe~\textsc{ii} lines 
at 2383.04\,\AA\ and 2388.63\,\AA\ depress the continuum
around the Te~\textsc{i} line.
}
\end{figure}

\begin{figure}
\begin{center}
\includegraphics[angle=0,width=3.0in]{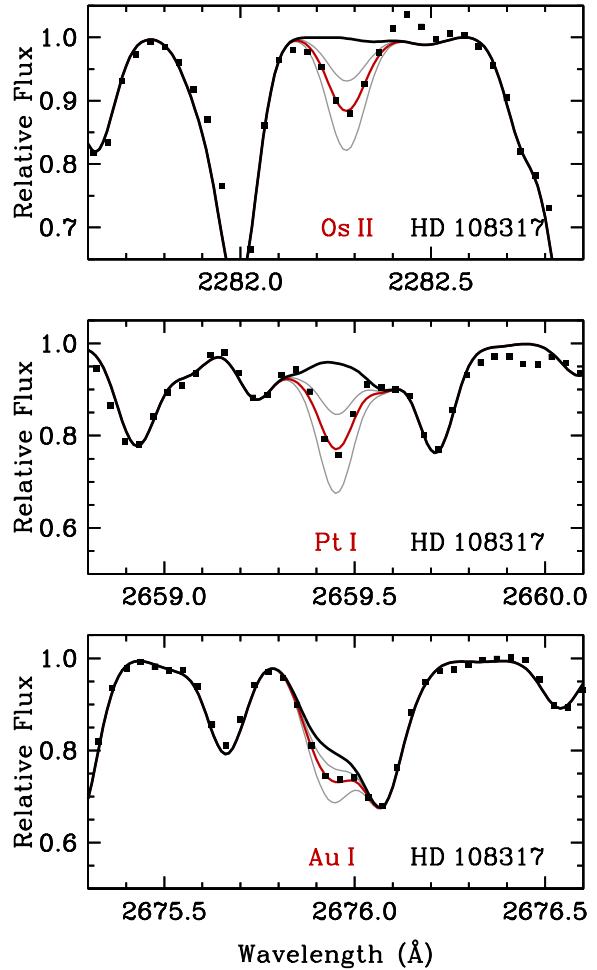}
\end{center}
\caption{
\label{synthplot3}
Comparison of observed and synthetic spectra in \hda.
The 
Os~\textsc{ii} 2282\,\AA,
Pt~\textsc{i} 2659\,\AA, and
Au~\textsc{i} 2675\,\AA\ lines are shown.
The bold red line represents the best-fit abundance,
the thin gray lines represent variations in this abundance by $\pm$~0.30~dex,
and the bold black line represents a synthesis with no
osmium, platinum, or gold.
}
\end{figure}

We have derived
abundances or upper limits for 
40~species of 37~elements heavier than zinc
in these four stars.
The final abundances and upper limits are reported in 
Tables~\ref{finalabund1} and \ref{finalabund2}.
The high S/N at short wavelengths has allowed us to 
derive abundances from larger numbers of lines than have been
useful previously
and detect
species whose only useful transitions are located shortward of 
$\approx$~2650\,\AA\
(e.g., Cd~\textsc{i}, Te~\textsc{i}, Lu~\textsc{ii}, Os~\textsc{ii}).
Several of our syntheses of important NUV transitions are 
illustrated in Figures~\ref{synthplot1}, \ref{synthplot2},
and \ref{synthplot3}.

As discussed in detail in Appendix~\ref{appendix},
we reconsider the cadmium abundance previously derived
from the lower S/N spectrum of \hdb\ examined by \citet{roederer10b}.
Cd~\textsc{i} is detected at 2288\,\AA,
but it is blended, and our synthesis 
of this spectral region is a poor match to the observed spectrum.
We report only an upper limit on the cadmium abundance in \hdb.

We detect neutral and singly-ionized species of 
osmium in \hda\ and \hdc.
These abundances are not in
agreement, with [Os~\textsc{i}/Fe] higher by 0.54 and 0.53~dex,
respectively.
\citet{roederer10b} reported a smaller (0.3~dex) discrepancy in \bd.
Adopting a different laboratory source for the Os~\textsc{ii} 2282\,\AA\ 
$\log(gf)$ value (\citealt{ivarsson04} rather than \citealt{quinet06}) 
would only decrease the discrepancy by 0.09~dex.
This offset is not due to 
missing energy levels in the 
Os~\textsc{i} and \textsc{ii} partition functions, and our tests
indicate it is not caused by neglecting isotope shifts in our synthesis.
The source of this discrepancy remains unclear.
Since only the neutral species of the neighboring elements
iridium, platinum, and gold are detected, 
and since most previous osmium abundances have been derived
using Os~\textsc{i},
we adopt Os~\textsc{i} as the primary osmium abundance indicator.

We are able to fit blends and derive more reliable
abundances of the third \rpro\ peak elements in \hda\
by assuming the third-peak elements in \hdd\ produce no absorption.
The gold abundance derived from the Au~\textsc{i} 2675\,\AA\ line
should be viewed cautiously given the strength of
the blending features
(Figure~\ref{synthplot3}).
We cannot exclude the possibility that a small amount of absorption
from Pt~\textsc{i} is present in \hdd.
Relaxing this assumption would change the platinum abundances
in \hda\ by $<+$0.1~dex.
Conservatively, we only report an upper limit on the platinum
abundance in \hdd.

We urge caution when interpreting abundances derived from
weak, blended, or small numbers of lines.
We have attempted to account for these factors in the
stated uncertainties, but unidentified blends 
may lead to spuriously high abundances.
The rhodium ($Z =$~45), 
silver ($Z =$~47),
praseodymium ($Z =$~59),
terbium ($Z =$~65), 
lutetium, 
hafnium,
osmium,
iridium, and
gold 
abundances are most susceptible to this bias.

Our abundances are not always in agreement with previous results for
\hdb.
To some extent this reflects our higher quality spectrum and the 
greater number of lines available to us now than in the past;
however, this explanation alone is insufficient.
For example,
\citet{cowan05} derived $\log\epsilon$~(Ge)~$= -$0.16 
from the Ge~\textsc{i} 3039\,\AA\ line in \hdb, 
while we derive
$\log\epsilon$~(Ge)~$= -$0.87.
We confirm that this discrepancy arises from
the updated version of MOOG used for the analysis.
(Rayleigh scattering has a $\lambda^{-4}$ dependence, so
differences between the two versions of MOOG
are most pronounced for
transitions at short wavelengths.)
Ratios constructed among the neutron-capture elements are in much better
agreement.
For example, 
after accounting for the different \loggf\ values
(all other atomic data---partition functions, etc.---are identical),
\citeauthor{cowan05}\ derived
[Zr/Ge]~$= +$1.12, while we derive
[Zr/Ge]~$= +$1.14.
This agreement is encouraging.
Nevertheless, the difference in $\log\epsilon$
illuminates the difficulty in obtaining reliable absolute abundances
and underscores our assertion that abundance ratios
are to be strongly preferred when interpreting the results.

\section{Discussion}
\label{discussion}

\begin{figure*}
\begin{center}
\includegraphics[angle=270,width=4.8in]{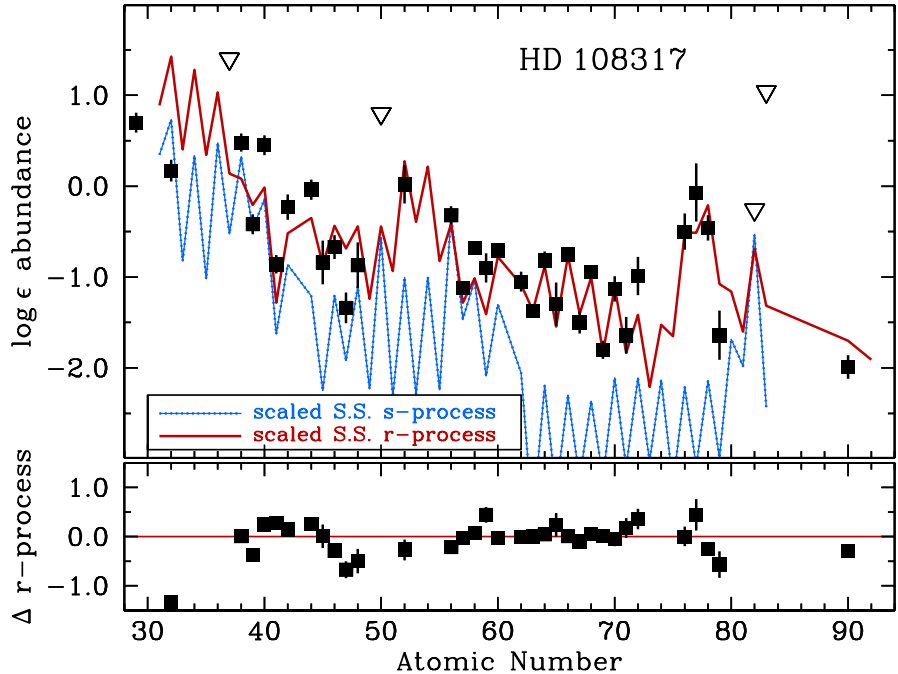}
\end{center}
\caption{
\label{loge108}
Comparison of the derived abundances in \hda\ with the 
scaled S.S.\ \spro\ (studded blue line, normalized to Ba) 
and \rpro\ (smooth red line, normalized to Eu) abundance predictions.
The top panel indicates the logarithmic abundances, and
the bottom panel indicates the residuals between the 
stellar abundances and the \rpro\ distribution
when normalized to Eu.
Filled squares indicate detections, and open downward facing triangles
indicate upper limits.
The Os~\textsc{ii} abundance has been
omitted in favor of Os~\textsc{i}.
We advise that ratios of elements derived from different ionization states
should be compared with caution.
}
\end{figure*}

\begin{figure*}
\begin{center}
\includegraphics[angle=270,width=4.8in]{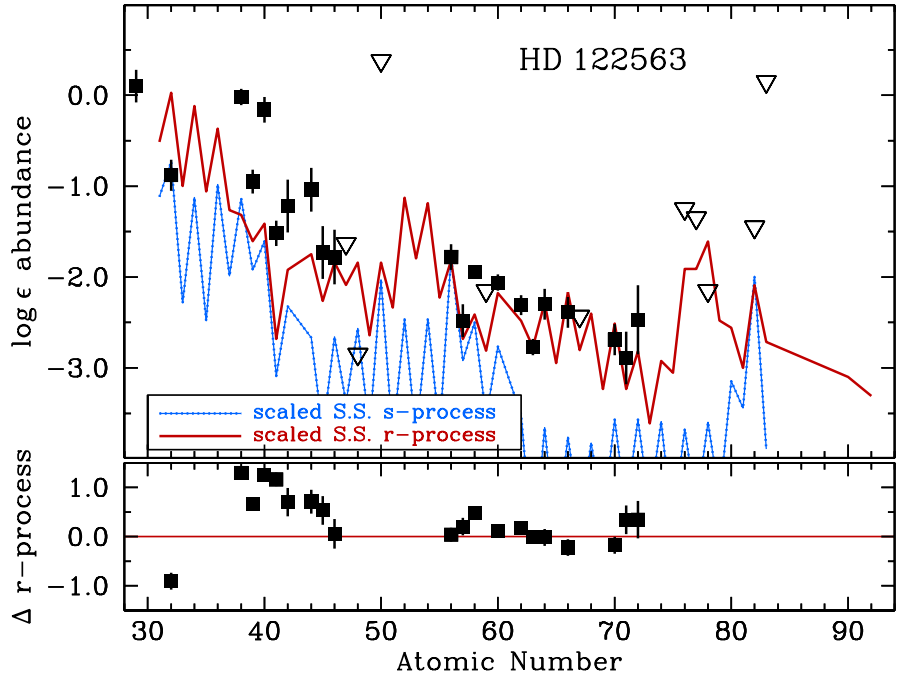}
\end{center}
\caption{
\label{loge122}
Comparison of the derived abundances in \hdb\ with the 
scaled S.S.\ \spro\ (studded blue line, normalized to Ba) 
and \rpro\ (smooth red line, normalized to Eu) abundance predictions.
The top panel indicates the logarithmic abundances, and
the bottom panel indicates the residuals between the 
stellar abundances and the \rpro\ distribution
when normalized to Eu.
Filled squares indicate detections, and open downward facing triangles
indicate upper limits.
\citet{honda06} detected silver in \hdb\ 
in their higher-quality spectrum.
When compared with the adjacent element palladium,
our upper limit is consistent with their detection.
We advise that ratios of elements derived from different ionization states
should be compared with caution.
}
\end{figure*}

\begin{figure*}
\begin{center}
\includegraphics[angle=270,width=4.8in]{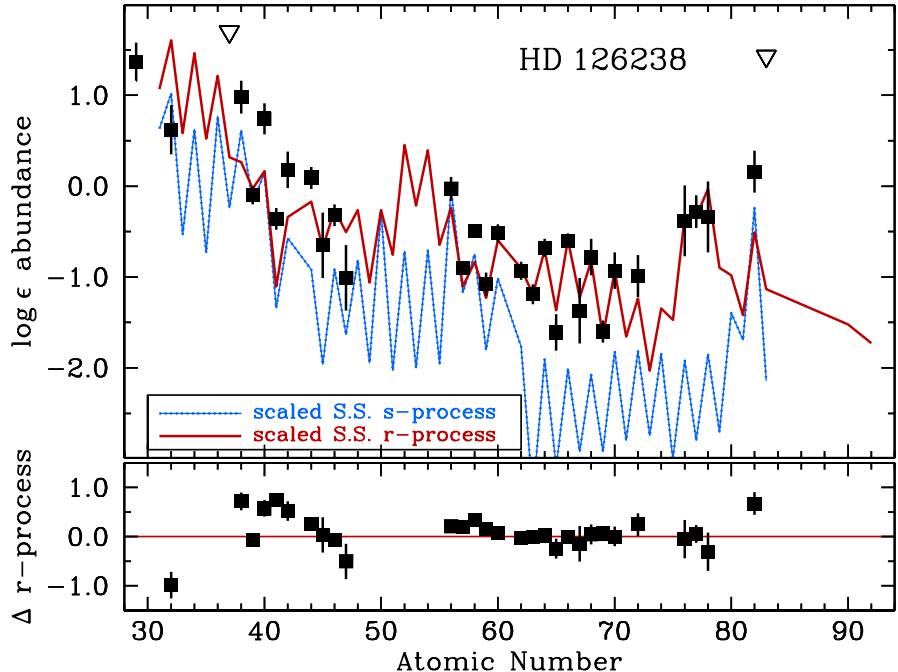}
\end{center}
\caption{
\label{loge126}
Comparison of the derived abundances in \hdc\ with the 
scaled S.S.\ \spro\ (studded blue line, normalized to Ba) 
and \rpro\ (smooth red line, normalized to Eu) abundance predictions.
The top panel indicates the logarithmic abundances, and
the bottom panel indicates the residuals between the 
stellar abundances and the \rpro\ distribution
when normalized to Eu.
Filled squares indicate detections, and open downward facing triangles
indicate upper limits.
The Os~\textsc{ii} abundance has been
omitted in favor of Os~\textsc{i}.
We advise that ratios of elements derived from different ionization states
should be compared with caution.
}
\end{figure*}

\begin{figure*}
\begin{center}
\includegraphics[angle=270,width=4.8in]{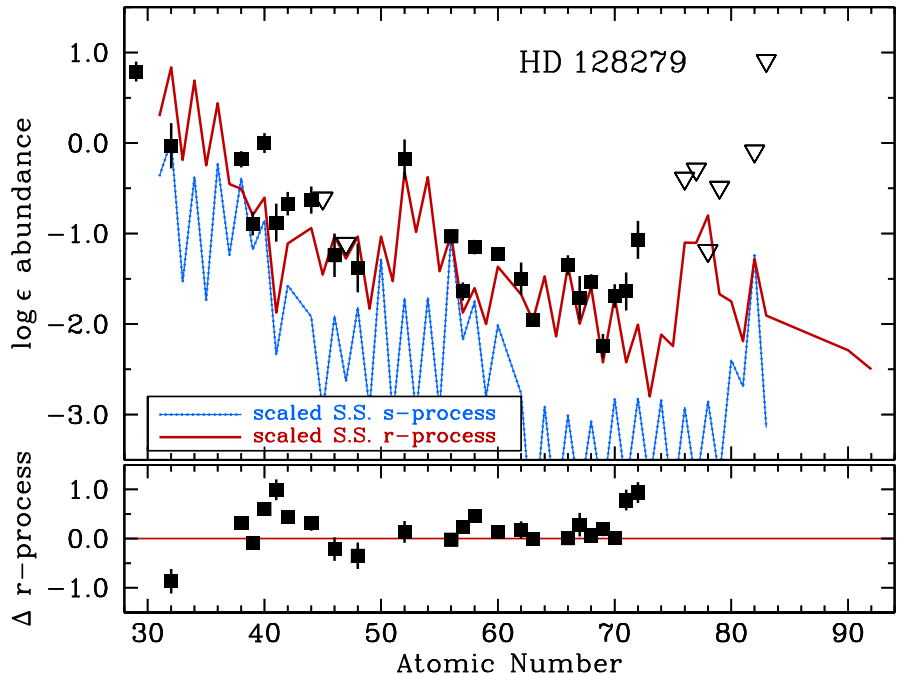}
\end{center}
\caption{
\label{loge128}
Comparison of the derived abundances in \hdd\ with the 
scaled S.S.\ \spro\ (studded blue line, normalized to Ba) 
and \rpro\ (smooth red line, normalized to Eu) abundance predictions.
The top panel indicates the logarithmic abundances, and
the bottom panel indicates the residuals between the 
stellar abundances and the \rpro\ distribution
when normalized to Eu.
Filled squares indicate detections, and open downward facing triangles
indicate upper limits.
We advise that ratios of elements derived from different ionization states
should be compared with caution.
}
\end{figure*}

Figures~\ref{loge108} through \ref{loge128}
illustrate the abundance pattern for each star.
In these figures we compare the $\log\epsilon$ abundances with the
predicted distributions of $s$- and \rpro\ material in the S.S., scaled
downward to match the stellar abundance levels.
The predicted $s$- and \rpro\ distributions shown 
in Figures~\ref{loge108} through \ref{loge128} 
are taken from \citet{sneden08}.
The lead and bismuth predictions are taken from the 
low-metallicity stellar models of \citet{bisterzo11}
that reproduce the ``strong'' component of S.S.\ lead and bismuth.
Recall that the S.S.\ \rpro\ distribution implicitly includes
contributions from all processes other than the \spro,
including an unspecified light element primary process 
(LEPP; \citealt{travaglio04}).

Figures~\ref{loge108} through \ref{loge128} reveal that the abundance
patterns for $Z \geq$~52
in \hda, \hdc, and \hdd\ (except for $Z >$~70 in \hdd) 
match the scaled S.S.\ \rpro\
distribution more closely than the \spro\ distribution.
The lighter elements also favor the \rpro\ distribution but with
greater element-to-element scatter than seen in the heavier elements.
Fitting the observed abundance distributions with a combination of
the S.S.\ \rpro\ abundance pattern and a small additional amount
of \spro\ material offers marginal improvement
for \hda\ and \hdc.
The lead abundances we have derived from an LTE analysis 
could be underestimated
by factors of $\approx$~2--3 \citep{mashonkina12}.
Consequently, according to \citet{roederer10c},
the enhanced lead abundance in \hdc\ could indicate a small 
\spro\ contribution.
The abundance pattern in \hdb\ clearly disfavors the \spro\ distribution,
but it is not as well-matched to the \rpro\ distribution as 
the other stars are.
No combination of $s$- and \rpro\ material can account for
the observed abundance distribution in \hdb, as has
been noted previously
\citep{sneden83,honda06}.
\hdd\ is a probable member of a stellar stream.
The more metal-rich stars in this stream do not show
evidence of \spro\ enrichment \citep{roederer10a}, so it would be surprising
if \hdd\ contained much \spro\ material.

We refrain from making any quantitative assessments 
of the \spro\ contamination. 
Fitting linear combinations of two distributions 
assumes that the S.S.\ distributions
are representative at low metallicity
and that no other processes contribute.
There is ample evidence (e.g., \citealt{wasserburg96};
\citealt{mcwilliam98})
that isotopes with $A \lesssim$~130 do not always
adhere to the scaled S.S.\ \rpro\ distribution when scaled
to heavier isotopes.
These elements, including strontium, yttrium, and zirconium,
may be produced by nucleosynthesis channels separate from---but
perhaps sometimes associated with---\rpro\ nucleosynthesis.
This complicates the interpretation of the \rpro\ residual
distribution for $A \lesssim$~130 isotopes.
There is no single distribution produced by all
\spro\ environments, and the resulting abundance ratios are known to vary
depending on the neutron density, availability of
$^{13}$C, timescales, number of dredge-up episodes, initial metallicity, 
etc.\ (e.g., \citealt{gallino98,bisterzo10}).
Variations observed in the [La/Eu] ratio in stars with very low
lead abundances ([Pb/Eu]~$< -$0.7) also may point to intrinsic
variations in \rpro\ distributions \citep{roederer10c}.
We cannot exclude the possibility of small amounts of \spro\ material
in \hda\ and \hdc, but clear evidence 
of \spro\ contamination is not apparent in \hdb\ or \hdd.
Either way, the \spro\ contamination to many of the
elements of interest in the present study 
is insignificant.

When detected, the ratios among 
palladium, silver, and cadmium
are relatively constant.
This agrees with the earlier results of \citet{johnson02},
\citet{roederer12b}, 
and a substantially larger sample of stars examined
by \citet{hansen11} and \citet{hansen12}.
The work of \citeauthor{hansen12}\
suggests that palladium and silver are
produced under \rpro\ conditions that are different---less
extreme neutron densities, perhaps---from 
the conditions that produce heavier mass nuclei,
such as europium ($Z =$~63), via the main component of the \rpro.
Our results, as well as those of \citet{roederer10b} and \citet{roederer12b},
imply that cadmium 
was likely produced along with palladium and silver
in the events that enriched \bd, \hda, \hdd, and \mbox{HD~160617}.

We recover the well-known
downward trends of abundance with increasing atomic number
between the first and second \rpro\ peaks 
in \hdb\ (e.g., \citealt{truran02,honda06,montes07}).
The contrast between \hdb\ and the stars discussed in the
previous paragraph is
illustrated in stark terms by the upper limit on the
cadmium abundance in \hdb\ (Figure~\ref{loge122}), which 
is more than one dex lower than the abundance of
palladium.
This indicates that production 
declined sharply through this mass range, yet
some form of \rpro\ nucleosynthesis 
seems to have produced the rare earth elements
with masses between 
$A \approx$~135 and 180.

The elements at the second and third \rpro\ peaks in \hda\ 
agree with the scaled S.S.\ \rpro\ distribution
within factors of $\approx$~2.
Similar results have been found for the metal-poor halo stars
\bd\ and \mbox{HD~160617} by \citet{roederer12a} and 
\citet{roederer12b}.
While the number of halo stars with elements detected at 
both the second and third \rpro\ peaks is still small,
these results hint that tellurium was mainly produced
along with the heavier \rpro\ elements in the
events that enriched these stars.

Upper limits on the platinum in \hdb\ and \hdd\
suggest that the third \rpro\ peak elements here are 
deficient relative to the scaled S.S.\ \rpro\ distribution.
Are these patterns
indicative of separate nucleosynthesis processes, or are they
the outcomes of high- and low-intensity 
\rpro\ nucleosynthesis?
\citet{kratz07} have used a set of site-independent waiting-point
\rpro\ calculations to show that the combined yields of
exposures with different neutron densities can reproduce the
overall shape of the
S.S.\ \rpro\ distribution from the first \rpro\ peak to the actinides.
Those calculations suggest that neutron densities lower than
$n_{n} \sim 10^{24}$--$10^{26}$ could 
under-produce the third peak elements.
On the other hand, \citet{travaglio04}, \citet{montes07}, and
\citet{qian07} have suggested that mixtures of two generic
nucleosynthesis events (a main \rpro\ and a LEPP)
can explain the stellar abundance patterns (see also \citealt{qian08}),
including the lack of a correlation between LEPP enhancement and
overall europium enrichment.
Our new abundance derivations, particularly our detections and upper limits
of elements at the second and third \rpro\ peaks,
may aid theoretical explorations of these issues.

\section{Summary} 
\label{conclusions}

We have detected up to 
34 elements produced
by neutron-capture reactions in each of
\hda, \hdb, \hdc, and \hdd.
The unique capability of high-resolution spectroscopy in the NUV
with \textit{HST} and STIS has enabled us to detect seven
of these 34~elements that cannot be detected from the ground.
\hda\ and the strongly \rpro-enriched star \bd, itself the subject
of numerous ground- and space-based abundance studies
(including, e.g., \citealt{burris00,cowan02,cowan05,denhartog05,
francois07,sneden09,roederer10b,roederer12a}), 
are the two metal-poor halo stars 
with the largest number of heavy elements detected.

We explore several potential sources of systematic uncertainties
in the abundance analysis.
Our \teff\ estimates, derived from
standard spectroscopic techniques, are consistently 
more than 200~K cooler than $V-K$ photometric estimates, 
and for \hdb\ our spectroscopic \teff\ is 150~K cooler 
than the \teff\ derived from a 
recent measurement of the stellar radius.
In general, abundance ratios are more reliable than absolute
abundances.
Ratios constructed among elements
whose abundances have been derived from species in the
same ionization state are more reliable
than ratios mixing neutral atoms and ions.
Our iron abundances reveal wavelength-dependent abundance trends
that cannot be explained by non-LTE effects and may
indicate shortcomings in the calculation of the continuous opacity.
We are continuing to investigate this matter.
We make empirical corrections for this effect, but we 
encourage those who wish to make use of our derived abundances
to exercise due respect for these systematic uncertainties.

We find that
some form of \rpro\ nucleosynthesis is responsible
for the heavy elements in these stars, though we cannot exclude
the possibility of a small trace of \spro\ material in
\hda\ and \hdc.
Our results, combined with previous results for \bd\ and 
\mbox{HD~160617},
reveal the \rpro\ nucleosynthesis patterns in six stars with \rpro\ 
enrichment levels spanning a range of a factor of 40:
[Eu/Fe]~$= +$0.9, $+$0.5, $+$0.4, $+$0.2, 0.0, and $-$0.7 in
\bd, \hda, \mbox{HD~160617}, \hdc, \hdd, and \hdb, respectively.
The enrichment levels of the lighter elements, like strontium,
yttrium, and zirconium,
vary only by factors of 2--8.
The relative heights of the second and third \rpro\ peaks are
broadly consistent with the scaled S.S.\ \rpro\ distribution in \bd,
\hda, and \mbox{HD~160617}.
There is a general decrease in abundance with atomic number in
\hdb, especially between the first and second \rpro\ peaks,
that differs from the other stars examined.

While our new observations are no doubt useful to
identify the source(s) of the heavy elements in these 4 stars,
their greater impact surely lies in their ability to 
inform models of \rpro\ nucleosynthesis.
Our work has demonstrated that several previously-undetected 
elements key to this understanding
can be detected and reliably measured in ancient halo stars.
We trust that these new observational constraints 
will be of use in theoretical explorations of \rpro\ nucleosynthesis.

\acknowledgments

We express sincere appreciation to
G.\ Preston for obtaining some of the MIKE observations,
C.J.\ Hansen for sending results in advance of publication,
K.-L.\ Kratz and F.\ Montes for helpful discussions,
and the anonymous referee for several suggestions that 
we believe have improved this paper significantly.
We also appreciate the expert assistance of the STScI staff
in obtaining these observations and the 
astronauts of STS-125 for their enthusiastic
return to \textit{HST} for Servicing Mission~4.

This research has made use of NASA's 
Astrophysics Data System Bibliographic Services, 
the arXiv pre-print server operated by Cornell University, 
the SIMBAD and VizieR databases hosted by the
Strasbourg Astronomical Data Center,
the Atomic Spectra Database hosted by
the National Institute of Standards and Technology, 
the Mikulski Archive at the Space Telescope Science Institute, and
the NUV and visible spectral atlases of Arcturus and the Sun
\citep{hinkle00,hinkle05}.
IRAF is distributed by the National Optical Astronomy Observatories,
which are operated by the Association of Universities for Research
in Astronomy, Inc., under cooperative agreement with the National
Science Foundation.
This publication makes use of data products from the Two Micron All Sky Survey, 
which is a joint project of the University of Massachusetts and the 
Infrared Processing and Analysis Center/California Institute of Technology, 
funded by the National Aeronautics and Space Administration and the 
National Science Foundation.
We recognize and acknowledge the very significant cultural 
role and reverence that the summit of Mauna Kea has always had within the 
indigenous Hawaiian community.  
We are most fortunate to have the opportunity to conduct observations 
from this mountain.

Generous support for Program number 12268 was provided by NASA through
a grant from the Space Telescope Science Institute, which is operated by the
Association of Universities for Research in Astronomy, Incorporated, under
NASA contract NAS~5-26555.
I.U.R.\ is supported by the Carnegie Institution for Science
through a Carnegie Fellowship.
J.E.L.\ acknowledges support from NASA Grant NNX10AN93G.~
T.C.B.\ and H.S.\ acknowledge partial support from grants PHY 02-16783 and
PHY 08-22648: Physics Frontier Center / Joint Institute for Nuclear
Astrophysics (JINA), awarded by the U.S.\ National Science Foundation
(NSF).~
H.S.\ acknowledges additional support from NSF Grant PHY-1102511.
C.S.\ acknowledges support from NSF Grant AST~09-08978.

 {\it Facilities:} 
\facility{HST (STIS)},
\facility{Keck~I (HIRES)},
\facility{Magellan:Clay (MIKE)}

\appendix
\section{Comments on NUV Line Selection and Oscillator Strengths}
\label{appendix}

Blending from atomic and molecular contaminants
must be carefully assessed 
to reliably identify absorption due to heavy elements in the
NUV spectra of metal-poor stars.
Very few transitions in the NUV are unblended, so our
abundance analysis must proceed by spectral synthesis.
Our initial synthesis line lists include all atomic and 
molecular (OH) transitions in the \citet{kurucz95} line lists.
We update the $\log(gf)$ values with experimental data when known.
Here, we discuss the heavy element lines of interest in our
STIS spectra and lines that blend with them, as revealed by the
observed line profiles.
The $\log(gf)$ values of blending features
come from the \citeauthor{kurucz95}\ lists
unless noted otherwise.
Confidence in the accuracy of these values should be tempered.

\textit{Zinc ($Z =$~30)}---Zinc is traditionally 
considered an iron group element, 
but neutron capture 
reactions must pass through zinc isotopes when running
from lighter, 
more abundant iron group elements to the heavy elements.
A few good Zn~\textsc{i} transitions exist in the optical spectral range.
We report an additional Zn~\textsc{i} transition at 3075.90\,\AA\
found in our STIS spectra.
This transition is blended with an Fe~\textsc{i} line at
3075.72\,\AA\ 
(\loggfalt~$= -$0.68)
and a weak V~\textsc{i} transition at 3075.93\,\AA\
(\loggfalt~$= -$0.62).
The abundance
derived from this Zn~\textsc{i} transition agrees well with the abundance
derived from the 4680\,\AA, 4722\,\AA, and 4810\,\AA\ transitions.

\textit{Germanium ($Z =$~32)}---The 
Ge~\textsc{i} transition at 2651.17\,\AA\ is situated between 
several manageable blending features, most notably an OH line at
2651.30\,\AA\
(\loggfalt~$= -$2.88).
These blends are too strong to derive a reliable Ge~\textsc{i}
abundance in \hdc.
A nearby, but weaker, Ge~\textsc{i} transition at 2651.57\,\AA\ is 
too blended for abundance work in all four~stars.
The Ge~\textsc{i} transition at 2691.36\,\AA\ is relatively clean, although
unidentified absorption lines
in each wing (2691.20\,\AA\ and 2691.48\,\AA, the former itself in the wing
of a strong Cr~\textsc{ii} line at 2691.04\,\AA;
\loggfalt~$= -$0.40) are present in all four stars.
This line gives abundance results lower by 0.1--0.5~dex
than all other Ge~\textsc{i} lines considered,
but we have no other
compelling reason to exclude it as an abundance indicator.
The Ge~\textsc{i} transition at 3039.07\,\AA\ has been used previously
for abundance work, and we can model the only minor blending feature,
a weak Fe~\textsc{i} line at 3038.98\,\AA\
with no $\log(gf)$ value
given in the NIST database.
(The \citealt{kurucz95} lists give \loggfalt~$= -$2.59, which we adjust
from $-$2.0 to $-$2.4 to fit the observed line profile.)
Moderately strong Ge~\textsc{i} absorption lines are also detected at
2591.17\,\AA\ and 2754.59\,\AA. 
Blends with OH and Fe~\textsc{i}, respectively,
in addition to severe line blanketing from a nearby strong Fe~\textsc{ii}
line at 2755.74\,\AA\ 
in \hdb\ and \hdc,
render them unusable for abundance work.

\textit{Zirconium ($Z =$~40)}---We have identified eight Zr~\textsc{ii}
transitions in our STIS spectra 
(in addition to another 30 or so in the optical regime)
that are reliable abundance indicators in these stars.
The 2567.64\,\AA\ transition is unblended except in \hdb, where 
line blanketing from the nearby strong Fe~\textsc{ii} 2566.91\,\AA\ transition
depresses the continuum.
The 2699.60\,\AA\ transition suffers mild blends in the wings that can
be fit in all stars in our sample except for \hdb, where
extra absorption is present on the red side of the line.
Our syntheses suggest that the 2700.13\,\AA\ transition contains
a mild ($<$~10\%) OH blend on the blue side of the line and
even weaker blends on the red side from Fe~\textsc{i} and \textsc{ii}
and Co~\textsc{i}.
The S/N and resolution of our spectra are sufficient to fit these blends.
The 2732.72\,\AA\ transition is clean in \hda\ and \hdd, but the placement
of the continuum is too uncertain in \hdb\ and \hdc\ to yield a 
reliable abundance.
The 2758.81\,\AA\ transition is largely unblended in \hda, \hdb, and \hdd.
Our syntheses suggest that 
it is significantly blended with the V~\textsc{ii} 2758.82\,\AA\ 
transition 
(\loggfalt~$= -$0.54 in the \citealt{kurucz95} lists) in \hdc,
where it introduces a slight asymmetry in the line profile.
We can find no laboratory $\log(gf)$ value for this transition.
This V~\textsc{ii} absorption is expected to be no more than a minor
contaminant to the Zr~\textsc{ii} line in the other stars.
The 2915.99\,\AA\ transition lies on the blue wing of a moderately strong 
Fe~\textsc{ii} transition at 2916.15\,\AA, which we can fit;
NIST quotes $\log(gf) = -$3.31 for this line with an
uncertainty of 50\%.
Our syntheses suggest that the Fe~\textsc{ii} 
line is also blended with OH at 2916.23\,\AA\
(\loggfalt~$= -$2.26).
We derive an abundance from this Zr~\textsc{ii} transition in all four stars.
The 3054.84\,\AA\ transition is blended with Co~\textsc{ii} 3054.72\,\AA\
(\loggfalt~$= -$2.74)
and OH 3054.97$+$3054.99\,\AA\ in each wing
(\loggfalt~$= -$2.72 and $-$3.12, respectively), but these blends
can be reasonably fit in all four stars.
The 3095.07\,\AA\ transition is unblended in all four stars.

The 2639.08\,\AA\ transition is unblended, but its observed wavelength
is approximately 0.01\,\AA--0.03\,\AA\ redder (i.e., to longer wavelengths)
than predicted in all four stars.
In \hdc, the most metal-rich star in our sample, 
this line yields an abundance higher by a factor of four than the mean
of all other Zr~\textsc{ii} lines.
There are no obvious iron group blending transitions in the NIST 
database or the \citet{kurucz95} lists, but due to the possibility of an 
unidentified blend we discard this line from consideration 
in all stars in the sample.
We also examined Zr~\textsc{ii} 
transitions at 2571.39\,\AA, 2571.46\,\AA, 2752.20\,\AA, 2968.96\,\AA,
2969.62\,\AA, 3030.92\,\AA, 3036.39\,\AA, and 3036.50\,\AA.
All were found to be too blended for use here.
The 3061.33\,\AA\ transition is unblended in these stars, 
but it was not covered in either of the 
\citet{ljung06} or \citet{malcheva06} laboratory studies.
The \citeauthor{kurucz95} database reports $\log(gf) = -$1.38.
Using the zirconium abundance derived from the other lines
in these stars, we determine an empirical $\log(gf)$ value for
the 3061.34\,\AA\ line of $-$1.24~$\pm$~0.12.
We do not use this line in our analysis.

We have performed a detailed comparison of the Zr~\textsc{ii}
$\log(gf)$ values presented by \citet{ljung06} and \citet{malcheva06}
for all of the transitions used in our analysis.
It is reassuring that
the mean zirconium
abundance derived for each star using different sets of $\log(gf)$ 
values varies by no more than 0.04~dex.
The standard deviations of these measurements are
0.13--0.22~dex.
These values are slightly larger than the
standard deviations of the differences in the $\log(gf)$ values
between the two studies,
0.07--0.18~dex depending on which set of stellar lines is considered.
Neither study covers all of the transitions we have examined, unfortunately.
\citeauthor{malcheva06}\ suggest that their transition probabilities are
accurate to within 20\%, while \citeauthor{ljung06}\
estimate uncertainties of 4\%--11\% for the transitions we have analyzed.
Based on this we adopt the \citeauthor{ljung06}\ transition probabilities
whenever possible.
Five transitions in our list have highly discrepant $\log(gf)$ values
(differences of 0.32--0.60~dex) between the two studies, 
and for these we choose the $\log(gf)$ value that yields a zirconium
abundance nearer the mean abundance derived from other Zr~\textsc{ii} 
transitions.
All four stars in our sample consistently point to the 
same preferred $\log(gf)$ values.
Four of the five transitions (2732.72\,\AA, 3344.79\,\AA, 3403.68\,\AA, 
and 3549.51\,\AA)
favor the \citeauthor{ljung06}\ values, and one (2699.60\,\AA) 
favors the \citeauthor{malcheva06}\ value.

\textit{Niobium ($Z =$~41)}---The
Nb~\textsc{ii} transition at 2950.88\,\AA\ shows absorption in both wings from
OH at 2950.76\,\AA\ and 2950.94\,\AA\
(\loggfalt~$= -$2.33 and $-$3.23, respectively).
Treating the overall OH abundance as a free parameter we can fit these 
blends, and we use this niobium abundance indicator in all four stars.
The 3028.44\,\AA\ transition suffers only a blend in the blue wing with
an unidentified absorption feature.
The Nb~\textsc{ii} absorption is too weak to detect in \hdd, but 
otherwise it gives results in good accordance with the 2950.88\,\AA\ transition.
The Nb~\textsc{ii} 2827.08\,\AA\ line is too blended in these stars.
We detect what appears to be Nb~\textsc{ii} absorption at 
2876.96$+$2877.04\,\AA,
but there are too many blending features to reliably identify the 
continuum or derive a meaningful niobium abundance.

\textit{Molybdenum ($Z =$~42)}---The 
line profile of the Mo~\textsc{ii} 2871.51\,\AA\ transition clearly indicates
a blend, which may be due to Fe~\textsc{i} 2871.48\,\AA.
Unfortunately, no experimental $\log(gf)$ value is known for this transition.
The \citet{kurucz95} lists give \loggf~$= -$2.78.
It is too blended to fit empirically.
We only derive an upper limit on the molybdenum abundance
from this Mo~\textsc{ii} line.

\textit{Cadmium ($Z =$~48)}---Our syntheses suggest that only 
one line blends with the Cd~\textsc{i} 2288.02\,\AA\ transition,
an Fe~\textsc{i} transition at 2288.04\,\AA\ present in the Kurucz
line lists (\loggfalt~$= -$3.58).
Our line list can provide a reasonable match to the observed spectrum
in \hda\ and \hdd.
We derive a cadmium abundance in these two stars
and estimate uncertainties by empirically varying the blending Fe~\textsc{i}
line as far as allowed by the line profile.
The lower S/N and higher line density in \hdc\ at 2288\,\AA\
prevent us from reliably identifying the continuum, unfortunately,
so we cannot derive a meaningful cadmium abundance in this star.
Our synthesis provides a poor match to the overall spectral region
surrounding the Cd~\textsc{i} line in \hdb, and the relatively
high line density in this region, due to the much cooler temperatures,
also makes it difficult to identify the continuum.
We reported a tentative detection of Cd~\textsc{i} from this transition
in \citet{roederer10b}, but our higher S/N data indicate that 
an upper limit may be more appropriate.

\textit{Tellurium ($Z =$~52)}---The Te~\textsc{i} 2385.79\,\AA\ transition
is weak but lies in a relatively clean spectral window.
Strong Fe~\textsc{ii} lines at 2383.06\,\AA\ 
(\loggfalt~$= -$1.29 according to NIST) and 2388.63\,\AA\ 
(\loggfalt~$= -$0.14 according to NIST) depress the 
continuum by about 10\% in \hda\ and \hdd, which is still manageable.
In \hdb\ and \hdc\ the continuum is depressed by 40\%--70\%, rendering
the Te~\textsc{i} line useless as an abundance indicator.
In \hda\ and \hdd, nearby Fe~\textsc{i} transitions at 
2385.59\,\AA\ 
(\loggfalt~$= -$2.95) and 2385.92\,\AA\
(\loggfalt~$= -$4.15)
can be fit without significantly 
disturbing the line profile of the Te~\textsc{i} transition.
The only other minor blending feature suggested by our syntheses is
a weak Cr~\textsc{i} line at 2385.72\,\AA.
The NIST database gives $\log(gf) = -$0.88 for this line with a stated
uncertainty of 50\%, and we fit the overall line profile while varying
the strength of this Cr~\textsc{i} line.
Even accounting for the sub-Solar [Cr/Fe] ratios, the line profile
still demands that the \loggf\ of this line should be weaker,
from $-$1.5 to $-$1.7 based on these observations.

\textit{Lutetium ($Z =$~71)}---The continuum surrounding 
the Lu~\textsc{ii} ground-state transition at 2615.42\,\AA\ is 
depressed by about 5\% by neighboring Fe~\textsc{ii} lines at
2613.82\,\AA\ 
(\loggfalt~$= -$0.36 according to NIST) and 2617.62\,\AA\ 
(\loggfalt~$= -$0.52 according to NIST) in \hda\ and \hdd.
The line profile also suggests minor blends from Co~\textsc{i} at 2615.33\,\AA\
(\loggfalt~$= -$0.99, which we adjust to $-$1.5 assuming a
Solar [Co/Fe] ratio)
and OH at 2615.50\,\AA\
(\loggfalt~$= -$3.90), which can be fit moderately well.
In \hdb\ the continuum is depressed by about 20\%, and we report a 
tentative lutetium abundance in this star.
The continuum is depressed by about 30\% in \hdc\ and the blending is 
more severe, so we are unable to derive an abundance here
even though the Lu~\textsc{ii} line is strong and easily detected.
We present the 
hfs pattern for the Lu 2615\,\AA\ transition in
Appendix~\ref{luhfs}.

\textit{Hafnium ($Z =$~72)}---The Hf~\textsc{ii} 2641.41\,\AA\
transition is detected in all four stars as a weak absorption asymmetry between 
weaker Fe~\textsc{i} transitions at 2641.30\,\AA\ 
(\loggfalt~$= -$2.47, according to the \citealt{kurucz95} lists) 
and 2641.49\,\AA\
(\loggfalt~$= -$1.67, according to the \citeauthor{kurucz95}\ lists).
Laboratory $\log(gf)$ values are not known for either of these blends,
but we can adjust them to produce reasonable fits
to the line profile
(from $-$2.47 to $-$3.0 for the former, and 
 from $-$1.9  to $-$2.1 for the latter).
The Hf~\textsc{ii} line is extremely weak in \hdb.
Including hafnium in our synthesis does improve the fit to
the absorption line profile at this wavelength, so we
report a tentative detection.

\textit{Osmium ($Z =$~76)}---The Os~\textsc{ii} transition 
at 2282.28\,\AA\ is unblended in
both \hda\ and \hdd, although we do not detect any absorption from
osmium in \hdd.
This transition is clearly detected and
still relatively unblended in \hdc, suffering only
a few minor blends, the most notable arising from
an unidentified absorption feature at 2282.15\,\AA.
The S/N is relatively low at 2282\,\AA, so we report a tentative 
osmium abundance in \hdc.
No absorption is apparent at 2282.28\,\AA\ in \hdb.
Our syntheses have difficulty reproducing the overall observed spectrum
in this region in \hdb\ (see discussion regarding Cd~\textsc{i}).
The Os~\textsc{i} transition at
3058.66\,\AA\ lies between two stronger lines of Fe~\textsc{i} at
3058.49\,\AA\ 
(\loggfalt~$= -$0.50, which we adjust from
$-$1.0 to $-$1.1) and 3059.09\,\AA\
(\loggfalt~$= -$0.66, which we adjust from
$-$0.95 to $-$1.1).
There is no obvious absorption feature at this wavelength in 
\hdd, but we can improve the fit slightly by adjusting the strength
of a very weak Fe~\textsc{i} line at 3058.70\,\AA\
(\loggfalt~$= -$3.51).
In \hda, where absorption from Os~\textsc{i} is detected, 
this Fe~\textsc{i} blend serves to reduce the fraction of the 
absorption due to Os~\textsc{i}.
The Os~\textsc{i} 3058\,\AA\ transition gives consistently higher
abundances than the Os~\textsc{ii} 2282\,\AA\ transition.
If we do not adjust the strength of the Fe~\textsc{i} blend, however,
our derived Os~\textsc{i} abundance would be even higher
in \hda.
We include this adjusted Fe~\textsc{i}
line in the syntheses for the remaining stars in our sample.
We derive an osmium abundance from the 3058\,\AA\ line in 
\hdc\ but not \hdb, where no absorption is detected.
We adopt the $\log(gf)$ value for the Os~\textsc{ii} 
2282\,\AA\ transition from
\citet{quinet06}, so that the Os~\textsc{i} and Os~\textsc{ii}
$\log(gf)$ values are drawn from a common source.
Adopting instead the $\log(gf)$ value from \citet{ivarsson04} would have 
increased the derived abundance by only 0.09~dex.

\textit{Iridium ($Z =$~77)}---Only the Ir~\textsc{i} transition at
2924.79\,\AA\ is strong enough to detect in \hda\ and \hdc.
No absorption is detected at this wavelength in \hdb\ and \hdd.
The Ir~\textsc{i} line is significantly blended with the red wing of a strong 
composite feature dominated by 
V~\textsc{ii} 2924.64\,\AA\
(\loggfalt~$= +$0.15), which can be fit well.
\hda\ shows extra absorption at the correct wavelength relative to \hdd\
if the absorption arises from Ir~\textsc{i}, which is encouraging, 
but our derived iridium abundances should be viewed with caution.

\textit{Platinum ($Z =$~78)}---The 
Pt~\textsc{i} transition at 2659.45\,\AA\ is blended with
several weaker features, including Cr~\textsc{ii} 2659.46\,\AA\ 
(\loggfalt~$= -$0.94, which we adjust to $-$0.3 assuming a Solar
[Cr/Fe] ratio) and 
OH 2659.58\,\AA\
(\loggfalt~$= -$3.04).
The strength of these lines can be fit assuming that none of the
absorption at 2659.45\,\AA\ in \hdd\ comes from Pt~\textsc{i}.
If the $\log(gf)$ of the Cr~\textsc{ii} line---which is only
found in the Kurucz lists---is assumed to
be significantly lower, then the absorption could be due to
Pt~\textsc{i} in \hdd.
The \hdb\ spectrum is consistent with no absorption from either
Cr~\textsc{ii} or Pt~\textsc{i} here, so the weak line in
\hdd\ might be due to Pt~\textsc{i}.
We conservatively adopt only an upper limit on platinum in \hdd.
This Pt~\textsc{i} line
is easily detected in \hdc, but the blending features are
far too strong in this star to derive a reliable platinum abundance.

The Pt~\textsc{i} transition at 2929.78\,\AA\ is blended with 
a Cr~\textsc{ii} transition at 2929.80\,\AA, for which no laboratory
$\log(gf)$ is available.
(The \citealt{kurucz95} lists give \loggfalt~$= -$0.03, which we adjust
to $-$0.8 assuming a Solar [Cr/Fe] ratio.)
We can fit the strength of this line in \hdd, assuming that none
of the absorption arises from Pt~\textsc{i}, and use that to derive
a platinum abundance in \hda\ and \hdc.
No absorption is detected here in \hdb.
The Pt~\textsc{i} 2646.88\,\AA\ transition is
blended with an unidentified absorption feature at 2646.85\,\AA\
present in all of our spectra, including \hdb\ and \hdd.
After accounting for this absorption in our line list, only a very small
amount of extra absorption is visible in \hda\ and \hdc\ that
presumably comes from Pt~\textsc{i}.
We discard this line from further consideration.
There is an unidentified absorption feature at the exact wavelength
of the Pt~\textsc{i} 2771.66\,\AA\ transition.
This absorption does not change strength between \hda\ and \hdd\ 
as all other lines of heavy elements do, so we assume it is
not due to Pt~\textsc{i}.
This line and the Pt~\textsc{i} 2650.85\,\AA\ lines
originate from different lower levels of 0.10~eV;
\citet{denhartog05} determined the $\log(gf)$ value of
the 2650.85\,\AA\ transition to be 0.13~dex higher than that of the
2771\,\AA\ transition.
We cannot detect the 2650\,\AA\ transition, so it is reasonable to conclude
that the absorption at 2771.66\,\AA\ is not due to Pt~\textsc{i}.
The Pt~\textsc{i} transition at 2997.96\,\AA\ is blended with strong
OH transitions at 2997.96\,\AA\ 
(\loggfalt~$= -$2.07) and 2998.02\,\AA\
(\loggfalt~$= -$2.26),
preventing us from deriving 
a platinum abundance from this transition in any star in the sample.

\textit{Gold ($Z =$~79)}---The Au~\textsc{i} 2675.94\,\AA\ transition
is detected in \hda\ as extra absorption relative to \hdd.
There are a number of blends to this line, including 
OH 2675.89\,\AA\
(\loggfalt~$= -$3.15, according to the \citealt{kurucz95} lists),
Co~\textsc{i} 2675.98\,\AA\
(\loggfalt~$= -$1.66, according to the \citeauthor{kurucz95}\ lists), and
Ti~\textsc{i} 2676.07\,\AA\
(\loggfalt~$= -$1.78, according to the \citeauthor{kurucz95}\ lists, 
which we adjust from $-$0.15 to 0.0 assuming a Solar [Ti/Fe] ratio).
Laboratory $\log(gf)$ values are lacking for these transitions.
If we assume all absorption in \hdd\ is due to these three species, 
we can empirically fit the line profile, attributing the
remaining absorption in \hda\ to Au~\textsc{i}.
We advise that the derived abundance should be interpreted with
due caution.
The $\log(gf)$ values from \citet{morton00} and \citet{fivet06}
for this transition agree within 0.01~dex.

\textit{Lead ($Z =$~82)}---The Pb~\textsc{i} 2833.03\,\AA\ line
is blended with an Fe~\textsc{ii} line at 2833.09\,\AA.
NIST reports $\log(gf) = -$0.48 for this line with an uncertainty of 25\%.
We fit this line profile well in \hdd\ assuming that none of
the absorption is due to Pb~\textsc{i}.
Applying this fit directly to \hda\ reveals that only a minimal 
amount of absorption at this wavelength may be due to Pb~\textsc{i}, so
we only derive an upper limit.
The hfs/IS pattern for this transition is given
in Appendix~\ref{pbhfs}
along with three other Pb~\textsc{i} lines commonly used 
for abundance analyses.

\textit{Other Elements Not Detected}---There 
is an unidentified absorption line at 2943.58\,\AA\ 
whose EWs vary from 20\,m\AA--70\,m\AA\ in our sample.
This is close to the predicted wavelength of a low-excitation
Ga~\textsc{i} (gallium, $Z =$~31) transition at 2943.63\,\AA.
We have searched for absorption from other zero- or low-excitation
Ga~\textsc{i} transitions at 2874.24\,\AA, 4032.98\,\AA, and 4172.06\,\AA\
that should be present if the absorption at 2943.58\,\AA\ is Ga~\textsc{i}.
We detect none of these lines.
Gallium has a low first ionization potential of 6.00~eV and 
should mostly be present in these stars as Ga~\textsc{ii}.
We conclude that the absorption is not due to Ga~\textsc{i}.
We also synthesize regions around a number of low-excitation 
W~\textsc{i} and \textsc{ii} (tungsten, $Z =$~74) lines,
adopting $\log(gf)$ values from \citet{kling99} and \citet{morton00}.
None of these lines yield credible detections or 
interesting upper limits.
We synthesize a region around the Hg~\textsc{i}
(mercury, $Z =$~80)
2536.52\,\AA\ line, using the $\log(gf)$ value given by
\citet{morton00};
unfortunately, this line is far too blended to yield an interesting
upper limit.
We derive an upper limit on the bismuth 
abundance from the the Bi~\textsc{i} transition
at 3024.64\,\AA, adopting the $\log(gf)$ value given in the NIST
database.

\section{New hfs Patterns for Lu~II}
\label{luhfs}

\input{tab11}

In Table~\ref{lutab} we present the hfs line component pattern
for the Lu~\textsc{ii} 2615\,\AA\ transition.
There are two stable isotopes of lutetium, $^{175}$Lu and $^{176}$Lu.
The $^{175}$Lu isotope is dominant (97.4\% of S.S.\ lutetium;
\citealt{bohlke05}).
The $^{176}$Lu isotope is blocked from \rpro\
production by the stable $^{176}$Yb isotope, so we expect that
the lutetium in our sample of stars is primarily $^{175}$Lu.
The odd-$Z$ isotope $^{175}$Lu has non-zero nuclear spin $I =$~7/2,
which gives rise to the hfs.
The format of Table~\ref{lutab} is the same as Table~14 of \citet{lawler09},
which contains hfs component patterns for several other
Lu~\textsc{ii} lines throughout the optical and NUV.
These data were used previously to derive
the Lu~\textsc{ii} abundance in \bd\ and \hdb\ \citep{roederer10b}.
The component positions are given relative to the
center-of-gravity wavenumbers and air wavelengths.
Strengths are normalized to sum to one.
The positions are computed from the hfs constants given in
Table~10 of \citet{sneden03}, the energy levels given in
Table~12 of \citet{lawler09}, and the standard index of air given in
\citet{peck72}.

\section{New hfs Patterns for Pb~I}
\label{pbhfs}

\input{tab12}

In Table~\ref{pbtab} we present the hfs/IS line component pattern for
four Pb~\textsc{i} lines commonly used for abundance analysis.
There are four stable isotopes of lead, 
$^{204}$Pb, 
$^{206}$Pb, 
$^{207}$Pb, and
$^{208}$Pb.
The odd-$Z$ isotope $^{207}$Pb has non-zero nuclear spin $I =$~1/2,
which gives rise to the hfs.
The component positions are given relative to the
center-of-gravity wavenumbers and air wavelengths.
Strengths are normalized to sum to one 
for a S.S.\ isotopic composition, where
the fraction of each isotope, $f$, is
$f_{204}/f_{206}/f_{207}/f_{208} =$ 0.014/0.241/0.221/0.524
\citep{bohlke05}.
Energy levels are adopted from \citet{wood68}.
The hfs A values are adopted from \citet{bouazza00} for all lines.
The IS are adopted from \citeauthor{bouazza00}\ for the 
3639\,\AA, 3683\,\AA, and 4057\,\AA\
lines and \citet{thompson83} for the 2833\,\AA\ line.
The standard index of air is given in \citet{peck72}.

%
%
%
%
%
%
%
%
%
%
%

\end{document}

%% file: tab1.tex
\begin{deluxetable*}{ccccc}
\tablecaption{Log of Observations
\label{stistab}}
\tablewidth{0pt}
\tabletypesize{\scriptsize}
\tablehead{
\colhead{Star} &
\colhead{Program ID} &
\colhead{Dataset} &
\colhead{Date (UT)} &
\colhead{Time (ks)}}
\startdata
\multicolumn{5}{c}{Observations with \textit{HST}/STIS} \\
\hline
HD~108317 & GO-12268 & OBJQ01010--50 & 2011 Jun 11 & 12.9 \\
HD~108317 & GO-12268 & OBJQ02010--50 & 2011 Jul 13 & 12.9 \\
HD~108317 & GO-12268 & OBJQ03010--50 & 2011 Jul 15 & 12.9 \\
HD~122563 &  GO-8111 & O5EL01010--40 & 1999 Jul 29 & 10.4 \\
HD~122563 & GO-12268 & OBJQ09010--40 & 2011 Jul 24 &  9.9 \\
HD~122563 & GO-12268 & OBJQ10010--40 & 2011 Aug 06 &  9.9 \\
HD~126238 & GO-12268 & OBJQ07010--50 & 2011 Jul 20 & 13.4 \\
HD~126238 & GO-12268 & OBJQ08010--50 & 2011 Jul 21 & 13.4 \\
HD~128279 & GO-12268 & OBJQ04010--50 & 2011 Aug 24 & 13.0 \\
HD~128279 & GO-12268 & OBJQ05010--50 & 2011 Aug 26 & 13.0 \\
HD~128279 & GO-12268 & OBJQ06010--50 & 2011 Sep 05 & 13.0 \\
\hline
\multicolumn{5}{c}{Observations with Magellan/MIKE} \\
\hline
HD~108317 & \nodata  & \nodata       & 2009 Feb 21 & 0.12 \\
HD~122563 & \nodata  & \nodata       & 2009 Feb 19 & 0.03 \\
HD~126238 & \nodata  & \nodata       & 2009 Sep 04 & 0.10 \\
HD~128279 & \nodata  & \nodata       & 2004 Jun 22 & 0.02 \\
HD~128279 & \nodata  & \nodata       & 2004 Jul 14 & 0.20 \\
HD~128279 & \nodata  & \nodata       & 2004 Jul 23 & 0.90 \\
HD~128279 & \nodata  & \nodata       & 2005 May 31 & 0.24 \\
\hline
\multicolumn{5}{c}{Observations with Keck/HIRES} \\
\hline
HD~108317 & \nodata  & \nodata       & 1999 Apr 21 & 3.30 \\
HD~122563 & \nodata  & \nodata       & 1999 Apr 21 & 0.60 \\
HD~126238 & \nodata  & \nodata       & 1999 Apr 21 & 2.04 \\
HD~128279 & \nodata  & \nodata       & 1999 Apr 22 & 2.40 \\
\enddata
\end{deluxetable*}

%% file: tab2.tex
\begin{deluxetable*}{cccccccccc}
\tablecaption{Continuum Signal-to-Noise Estimates
\label{sntab}}
\tablewidth{0pt}
\tabletypesize{\scriptsize}
\tablehead{
\colhead{Star} &
\colhead{S/N} &
\colhead{S/N} &
\colhead{S/N} &
\colhead{S/N} &
\colhead{S/N} &
\colhead{S/N} &
\colhead{S/N} &
\colhead{S/N} &
\colhead{S/N} \\
\colhead{} &
\colhead{2290\AA} &
\colhead{2430\AA} &
\colhead{2650\AA} &
\colhead{3000\AA} &
\colhead{3520\AA} &
\colhead{4000\AA} &
\colhead{4550\AA} &
\colhead{5200\AA} &
\colhead{6750\AA} \\
\colhead{} &
\colhead{(STIS)} &
\colhead{(STIS)} &
\colhead{(STIS)} &
\colhead{(STIS)} &
\colhead{(HIRES)} &
\colhead{(HIRES)} &
\colhead{(MIKE)} &
\colhead{(MIKE)} &
\colhead{(MIKE)} }
\startdata
HD~108317 & 75/1 & 110/1 & 160/1 & 160/1 & 160/1 & 570/1 & 130/1 & 220/1 & 390/1 \\
HD~122563 & 60/1 & 120/1 & 260/1 & 400/1 & 220/1 & 640/1 & 160/1 & 290/1 & 540/1 \\
HD~126238 & 25/1 &  45/1 & 100/1 & 130/1 & 160/1 & 460/1 & 270/1 & 280/1 & 500/1 \\
HD~128279 & 75/1 & 110/1 & 160/1 & 180/1 & 140/1 & 350/1 & 880/1 & 500/1 & 750/1 \\
\enddata
\end{deluxetable*}

%% file: tab3-fauxstub.tex
\begin{deluxetable*}{ccccccccc}
\tablecaption{Atomic Data and Line-by-line Abundances
\label{atomictab}}
\tablewidth{0pt}
\tabletypesize{\scriptsize}
\tablehead{
\colhead{Species} &
\colhead{$\lambda$} &
\colhead{E.P.} &
\colhead{$\log(gf)$} &
\colhead{Ref.} &
\colhead{$\log \epsilon$} &
\colhead{$\log \epsilon$} &
\colhead{$\log \epsilon$} &
\colhead{$\log \epsilon$} \\
\colhead{} &
\colhead{(\AA)} &
\colhead{(eV)} &
\colhead{} &
\colhead{} &
\colhead{HD~108317} &
\colhead{HD~122563} &
\colhead{HD~126238} &
\colhead{HD~128279} 
}
\startdata
Cu~\textsc{i}   & 3247.54  &  0.00  &  $-$0.06  &   1  &   $+$0.67   &   \nodata  &   \nodata  &   $+$0.85  \\
Cu~\textsc{i}   & 3273.96  &  0.00  &  $-$0.36  &   1  &   $+$0.73   &   $+$0.10  &   $+$1.53  &   $+$0.73  \\
Cu~\textsc{i}   & 5105.54  &  1.39  &  $-$1.50  &   2  &  $<+$1.38   &  $<+$0.48  &   $+$1.21  &  $<+$1.08  \\
\enddata
\tablerefs{
(1) \citealt{fuhr09}, using hfs/IS from J.S.\ Sobeck et al.\ in prep.;
(2) \citealt{fuhr09}, using hfs/IS from \citealt{kurucz95};
(3) \citealt{roederer12b};
(4) \citealt{fuhr09};
(5) \citealt{migdalek87};
(6) \citealt{biemont11};
(7) \citealt{ljung06};
(8) \citealt{malcheva06};
(9) \citealt{nilsson08}, using hfs from \citealt{nilsson10} when available;
(10) \citealt{whaling88};
(11) \citealt{sikstrom01};
(12) \citealt{wickliffe94};
(13) \citealt{duquette85};
(14) \citealt{kwiatkowski82};
(15) \citealt{xu06};
(16) \citealt{hansen12} for both $\log(gf)$ and hfs/IS;
(17) \citealt{morton00};
(18) \citealt{roederer12a};
(19) \citealt{fuhr09}, using hfs/IS from \citealt{mcwilliam98};
(20) \citealt{lawler01a}, using hfs from \citealt{ivans06} when available;
(21) \citealt{lawler09};
(22) \citealt{li07}, using hfs from \citealt{sneden09};
(23) \citealt{denhartog03}, using hfs/IS from \citealt{roederer08} when available;
(24) \citealt{lawler06}, using hfs/IS from \citealt{roederer08} when available;
(25) \citealt{lawler01b}, using hfs/IS from \citealt{ivans06};
(26) \citealt{denhartog06};
(27) previously unpublished (uncertainty of 25\%);
(28) \citealt{lawler01c}, using hfs from \citealt{lawler09};
(29) \citealt{wickliffe00};
(30) \citealt{lawler04} for both $\log(gf)$ and hfs;
(31) \citealt{lawler08};
(32) \citealt{wickliffe97};
(33) \citealt{sneden09} for both $\log(gf)$ and hfs/IS;
(34) \citealt{roederer10b}, using hfs presented in Appendix~\ref{luhfs};
(35) \citealt{lawler07};
(36) \citealt{quinet06}, which is only 0.01~dex different from \citealt{cowan05} for this line;
(37) \citealt{quinet06};
(38) \citealt{xu07}, using hfs/IS from \citealt{cowan05} when available;
(39) \citealt{denhartog05} for both $\log(gf)$ and hfs/IS;
(40) \citealt{fivet06};
(41) \citealt{biemont00}, using hfs presented in Appendix~\ref{pbhfs};
(42) \citealt{nilsson02}.
}
\tablecomments{
The complete version of Table~\ref{atomictab} will be available in the 
final journal article.
The $\log\epsilon$ abundances listed in the Table have been corrected
according to the values given in Table~\ref{corrtab}.
 }
\tablenotetext{a}{
This feature is comprised of two Pr~\textsc{ii} transitions,
the 4429.13\,\AA\ transition with E.P.~$=$~0.00~eV and \loggf~$= -$0.49 and
the 4429.26\,\AA\ transition with E.P.~$=$~0.37~eV and \loggf~$= -$0.05.
 }
\end{deluxetable*}

%% file: tab4-stub.tex
\begin{deluxetable}{ccccc}
\tablecaption{Iron Equivalent Widths
\label{fetab}}
\tablewidth{0pt}
\tabletypesize{\scriptsize}
\tablehead{
\colhead{$\lambda$} &
\colhead{HD~108317} &
\colhead{HD~122563} &
\colhead{HD~126238} &
\colhead{HD~128279} \\
\colhead{(\AA)} &
\colhead{EW (m\AA)} &
\colhead{EW (m\AA)} &
\colhead{EW (m\AA)} &
\colhead{EW (m\AA)} 
}
\startdata
\multicolumn{5}{c}{Fe~\textsc{i}} \\
\hline
 3765.54 & \nodata & \nodata &   117.1 & \nodata \\
\enddata
\tablecomments{
The complete version of Table~\ref{fetab} is available online only.
A short version is shown here to demonstrate its form and content.
}
\end{deluxetable}

%% file: tab5.tex
\begin{deluxetable*}{ccccccccc}
\tablecaption{Basic Data
and Initial \teff\ and \logg\ Estimates
\label{datatab}}
\tablewidth{0pt}
\tabletypesize{\scriptsize}
\tablehead{
\colhead{Star} &
\colhead{$\pi$} &
\colhead{D} &
\colhead{$V$} &
\colhead{$K$} &
\colhead{$E(B-V)$} &
\colhead{$(V-K)_{0}$} &
\colhead{\teff$(V-K)$} &
\colhead{\logg$(\pi)$} \\
\colhead{} &
\colhead{(mas)} &
\colhead{(pc)} &
\colhead{} &
\colhead{} &
\colhead{} &
\colhead{} &
\colhead{(K)} &
\colhead{} 
}
\startdata
HD~108317 & 5.73$\pm$0.67 & 175$\pm$21 & 8.05 & 6.153 & 0.018 & 1.849 & 5310$\pm$75 & 2.90$\pm$0.16 \\
HD~122563 & 4.22$\pm$0.35 & 237$\pm$20 & 6.20 & 3.731 & 0.025 & 2.386 & 4680$\pm$65 & 1.58$\pm$0.13 \\
HD~126238 & 3.82$\pm$0.77 & 262$\pm$55 & 7.66 & 5.338 & 0.115 & 2.003 & 5100$\pm$72 & 2.09$\pm$0.26 \\
HD~128279 & 6.09$\pm$1.08 & 164$\pm$30 & 8.02 & 6.065 & 0.100 & 1.681 & 5550$\pm$79 & 2.85$\pm$0.23 \\
\enddata
\end{deluxetable*}

%% file: tab6.tex
\begin{deluxetable*}{ccccccccc}
\tablecaption{Adopted Model Parameters and Derived Metallicities
\label{modeltab}}
\tablewidth{0pt}
\tabletypesize{\scriptsize}
\tablehead{
\colhead{Star} &
\colhead{\teff} &
\colhead{\logg} &
\colhead{\vt} &
\colhead{[M/H]} &
\colhead{[Fe\,\textsc{i}/H]} &
\colhead{$N$} &
\colhead{[Fe\,\textsc{ii}/H]} &
\colhead{$N$} \\
\colhead{} &
\colhead{(K)} &
\colhead{} &
\colhead{(\kmsec)} &
\colhead{} &
\colhead{} &
\colhead{} &
\colhead{} &
\colhead{} 
}
\startdata
HD~108317 & 5100 & 2.67 & 1.50 & $-$2.37 & $-$2.53$\pm$0.09 & 300 & $-$2.37$\pm$0.14 & 55 \\
HD~122563 & 4450 & 1.37 & 2.00 & $-$2.61 & $-$3.06$\pm$0.10 & 211 & $-$2.61$\pm$0.12 & 20 \\
HD~126238 & 4780 & 1.72 & 1.60 & $-$1.93 & $-$1.98$\pm$0.10 & 217 & $-$1.93$\pm$0.10 & 26 \\
HD~128279 & 5080 & 2.57 & 1.60 & $-$2.46 & $-$2.48$\pm$0.09 & 266 & $-$2.46$\pm$0.14 & 48 \\
\enddata
\tablecomments{
The [Fe/H] values presented in this Table are derived from
individual line abundances that have been corrected 
according to the values presented inTable~\ref{corrtab}.
 }
\end{deluxetable*}

%% file: tab7.tex
\begin{deluxetable*}{ccccccccccccccc}
\tablecaption{Derived Final Abundances I
\label{finalabund1}}
\tablewidth{0pt}
\tabletypesize{\scriptsize}
\tablehead{
\colhead{} &
\colhead{} &
\colhead{S.S.} & 
\multicolumn{5}{c}{HD~108317} &
\colhead{} &
\multicolumn{5}{c}{HD~122563} \\
\cline{4-8} \cline{10-14} 
\colhead{Species} &
\colhead{$Z$} &
\colhead{$\log \epsilon$} &
\colhead{$\log \epsilon$} &
\colhead{[X/Fe]} &
\colhead{$\sigma_{\rm stat}$} &
\colhead{$\sigma_{\rm tot}$} &
\colhead{$N$} &
\colhead{} &
\colhead{$\log \epsilon$} &
\colhead{[X/Fe]} &
\colhead{$\sigma_{\rm stat}$} &
\colhead{$\sigma_{\rm tot}$} &
\colhead{$N$} 
}
\startdata
Fe~\textsc{i}  & 26 & $+$7.50 &  $+$4.97 &  $+$0.00 & 0.09    & 0.15    & 300     & &  $+$4.44 &  $+$0.00 & 0.10    & 0.16    & 211      \\
Fe~\textsc{ii} & 26 & $+$7.50 &  $+$5.13 &  $+$0.00 & 0.14    & 0.22    & 55      & &  $+$4.89 &  $+$0.00 & 0.12    & 0.20    & 20       \\
Cu~\textsc{i}  & 29 & $+$4.19 &  $+$0.70 &  $-$0.96 & 0.11    & 0.16    & 2       & &  $+$0.10 &  $-$1.03 & 0.18    & 0.22    & 1        \\
Zn~\textsc{i}  & 30 & $+$4.56 &  $+$2.26 &  $+$0.23 & 0.09    & 0.15    & 3       & &  $+$1.99 &  $+$0.49 & 0.17    & 0.21    & 4        \\
Ge~\textsc{i}  & 32 & $+$3.65 &  $+$0.17 &  $-$0.95 & 0.12    & 0.17    & 3       & &  $-$0.88 &  $-$1.47 & 0.17    & 0.21    & 3        \\
Rb~\textsc{i}  & 37 & $+$2.52 & $<+$1.40 & $<+$1.41 & \nodata & \nodata & 1       & & $<+$1.20 & $<+$1.74 & \nodata & \nodata & 1        \\
Sr~\textsc{ii} & 38 & $+$2.87 &  $+$0.48 &  $-$0.02 & 0.10    & 0.19    & 2       & &  $-$0.02 &  $-$0.28 & 0.09    & 0.19    & 3        \\
Y~\textsc{ii}  & 39 & $+$2.21 &  $-$0.42 &  $-$0.26 & 0.11    & 0.20    & 10      & &  $-$0.95 &  $-$0.55 & 0.13    & 0.21    & 10       \\
Zr~\textsc{ii} & 40 & $+$2.58 &  $+$0.45 &  $+$0.24 & 0.11    & 0.20    & 39      & &  $-$0.16 &  $-$0.13 & 0.14    & 0.22    & 26       \\
Nb~\textsc{ii} & 41 & $+$1.46 &  $-$0.86 &  $+$0.05 & 0.10    & 0.19    & 3       & &  $-$1.52 &  $-$0.37 & 0.14    & 0.22    & 3        \\
Mo~\textsc{i}  & 42 & $+$1.88 &  $-$0.23 &  $+$0.42 & 0.14    & 0.19    & 1       & &  $-$1.22 &  $-$0.04 & 0.29    & 0.31    & 1        \\
Mo~\textsc{ii} & 42 & $+$1.88 & $<+$0.24 & $<+$0.73 & \nodata & \nodata & 1       & & $<-$0.65 & $<+$0.08 & \nodata & \nodata & 1        \\
Ru~\textsc{i}  & 44 & $+$1.75 &  $-$0.04 &  $+$0.74 & 0.11    & 0.16    & 2       & &  $-$1.04 &  $+$0.27 & 0.24    & 0.27    & 3        \\
Rh~\textsc{i}  & 45 & $+$1.06 &  $-$0.84 &  $+$0.63 & 0.24    & 0.27    & 1       & &  $-$1.73 &  $+$0.27 & 0.29    & 0.31    & 1        \\
Pd~\textsc{i}  & 46 & $+$1.65 &  $-$0.67 &  $+$0.21 & 0.13    & 0.18    & 2       & &  $-$1.78 &  $-$0.37 & 0.30    & 0.32    & 1        \\
Ag~\textsc{i}  & 47 & $+$1.20 &  $-$1.34 &  $-$0.01 & 0.17    & 0.21    & 2       & & $<-$1.63 & $<+$0.23 & \nodata & \nodata & 2        \\
Cd~\textsc{i}  & 48 & $+$1.71 &  $-$0.87 &  $-$0.05 & 0.25    & 0.28    & 1       & & $<-$2.85 & $<-$1.50 & \nodata & \nodata & 1        \\
Sn~\textsc{i}  & 50 & $+$2.07 & $<+$0.80 & $<+$1.26 & \nodata & \nodata & 1       & & $<+$0.38 & $<+$1.37 & \nodata & \nodata & 1        \\
Te~\textsc{i}  & 52 & $+$2.18 &  $+$0.02 &  $+$0.37 & 0.21    & 0.24    & 1       & &  \nodata &  \nodata & \nodata & \nodata & \nodata  \\
Ba~\textsc{ii} & 56 & $+$2.18 &  $-$0.32 &  $-$0.13 & 0.10    & 0.19    & 4       & &  $-$1.78 &  $-$1.35 & 0.14    & 0.22    & 2        \\
La~\textsc{ii} & 57 & $+$1.10 &  $-$1.12 &  $+$0.15 & 0.05    & 0.17    & 7       & &  $-$2.48 &  $-$0.97 & 0.18    & 0.24    & 3        \\
Ce~\textsc{ii} & 58 & $+$1.58 &  $-$0.68 &  $+$0.11 & 0.04    & 0.17    & 11      & &  $-$1.94 &  $-$0.91 & 0.07    & 0.18    & 6        \\
Pr~\textsc{ii} & 59 & $+$0.72 &  $-$0.90 &  $+$0.75 & 0.16    & 0.23    & 2       & & $<-$2.15 & $<-$0.26 & \nodata & \nodata & 1        \\
Nd~\textsc{ii} & 60 & $+$1.42 &  $-$0.71 &  $+$0.24 & 0.07    & 0.18    & 12      & &  $-$2.06 &  $-$0.87 & 0.09    & 0.19    & 3        \\
Sm~\textsc{ii} & 62 & $+$0.96 &  $-$1.05 &  $+$0.36 & 0.11    & 0.20    & 5       & &  $-$2.31 &  $-$0.66 & 0.11    & 0.20    & 3        \\
Eu~\textsc{ii} & 63 & $+$0.52 &  $-$1.37 &  $+$0.48 & 0.06    & 0.18    & 5       & &  $-$2.77 &  $-$0.68 & 0.09    & 0.19    & 3        \\
Gd~\textsc{ii} & 64 & $+$1.07 &  $-$0.82 &  $+$0.48 & 0.10    & 0.19    & 6       & &  $-$2.30 &  $-$0.76 & 0.17    & 0.24    & 1        \\
Tb~\textsc{ii} & 65 & $+$0.30 &  $-$1.30 &  $+$0.77 & 0.24    & 0.29    & 1       & &  \nodata &  \nodata & \nodata & \nodata & \nodata  \\
Dy~\textsc{ii} & 66 & $+$1.10 &  $-$0.75 &  $+$0.52 & 0.08    & 0.18    & 9       & &  $-$2.39 &  $-$0.88 & 0.17    & 0.24    & 1        \\
Ho~\textsc{ii} & 67 & $+$0.48 &  $-$1.50 &  $+$0.39 & 0.12    & 0.20    & 2       & & $<-$2.43 & $<-$0.30 & \nodata & \nodata & 1        \\
Er~\textsc{ii} & 68 & $+$0.92 &  $-$0.94 &  $+$0.51 & 0.06    & 0.18    & 5       & &  \nodata &  \nodata & \nodata & \nodata & \nodata  \\
Tm~\textsc{ii} & 69 & $+$0.10 &  $-$1.80 &  $+$0.47 & 0.10    & 0.19    & 2       & &  \nodata &  \nodata & \nodata & \nodata & \nodata  \\
Yb~\textsc{ii} & 70 & $+$0.92 &  $-$1.13 &  $+$0.32 & 0.14    & 0.22    & 1       & &  $-$2.69 &  $-$1.00 & 0.17    & 0.24    & 1        \\
Lu~\textsc{ii} & 71 & $+$0.10 &  $-$1.64 &  $+$0.63 & 0.20    & 0.26    & 1       & &  $-$2.89 &  $-$0.38 & 0.29    & 0.33    & 1        \\
Hf~\textsc{ii} & 72 & $+$0.85 &  $-$0.99 &  $+$0.53 & 0.21    & 0.27    & 2       & &  $-$2.47 &  $-$0.71 & 0.38    & 0.41    & 1        \\
Os~\textsc{i}  & 76 & $+$1.40 &  $-$0.50 &  $+$0.63 & 0.20    & 0.23    & 1       & & $<-$1.25 & $<+$0.41 & \nodata & \nodata & 1        \\
Os~\textsc{ii} & 76 & $+$1.40 &  $-$0.88 &  $+$0.09 & 0.20    & 0.26    & 1       & &  \nodata &  \nodata & \nodata & \nodata & \nodata  \\
Ir~\textsc{i}  & 77 & $+$1.38 &  $-$0.07 &  $+$1.08 & 0.32    & 0.34    & 3       & & $<-$1.35 & $<+$0.33 & \nodata & \nodata & 3        \\
Pt~\textsc{i}  & 78 & $+$1.62 &  $-$0.46 &  $+$0.45 & 0.14    & 0.19    & 2       & & $<-$2.15 & $<-$0.71 & \nodata & \nodata & 2        \\
Au~\textsc{i}  & 79 & $+$0.80 &  $-$1.64 &  $+$0.09 & 0.27    & 0.30    & 1       & &  \nodata &  \nodata & \nodata & \nodata & \nodata  \\
Pb~\textsc{i}  & 82 & $+$2.04 & $<-$0.26 & $<+$0.23 & \nodata & \nodata & 3       & & $<-$1.45 & $<-$0.43 & \nodata & \nodata & 3        \\
Bi~\textsc{i}  & 83 & $+$0.65 & $<+$1.04 & $<+$2.92 & \nodata & \nodata & 1       & & $<+$0.15 & $<+$2.56 & \nodata & \nodata & 1        \\
Th~\textsc{ii} & 90 & $+$0.06 &  $-$1.99 &  $+$0.32 & 0.13    & 0.21    & 1       & &  \nodata &  \nodata & \nodata & \nodata & \nodata  \\
\enddata
\end{deluxetable*}

%% file: tab8.tex
\begin{deluxetable*}{ccccccccccccccc}
\tablecaption{Derived Final Abundances II
\label{finalabund2}}
\tablewidth{0pt}
\tabletypesize{\scriptsize}
\tablehead{
\colhead{} &
\colhead{} &
\colhead{S.S.} & 
\multicolumn{5}{c}{HD~126238} &
\colhead{} &
\multicolumn{5}{c}{HD~128279} \\
\cline{4-8} \cline{10-14} 
\colhead{Species} &
\colhead{$Z$} &
\colhead{$\log \epsilon$} &
\colhead{$\log \epsilon$} &
\colhead{[X/Fe]} &
\colhead{$\sigma_{\rm stat}$} &
\colhead{$\sigma_{\rm tot}$} &
\colhead{$N$} &
\colhead{} &
\colhead{$\log \epsilon$} &
\colhead{[X/Fe]} &
\colhead{$\sigma_{\rm stat}$} &
\colhead{$\sigma_{\rm tot}$} &
\colhead{$N$} 
}
\startdata
Fe~\textsc{i}  & 26 & $+$7.50 &  $+$5.52 &  $+$0.00 & 0.10    & 0.16    & 217     & &  $+$5.02 &  $+$0.00 & 0.09    & 0.15    & 266     \\
Fe~\textsc{ii} & 26 & $+$7.50 &  $+$5.57 &  $+$0.00 & 0.10    & 0.19    & 26      & &  $+$5.04 &  $+$0.00 & 0.14    & 0.22    & 48      \\
Cu~\textsc{i}  & 29 & $+$4.19 &  $+$1.37 &  $-$0.84 & 0.21    & 0.24    & 2       & &  $+$0.79 &  $-$0.92 & 0.11    & 0.16    & 2       \\
Zn~\textsc{i}  & 30 & $+$4.56 &  $+$2.64 &  $+$0.06 & 0.10    & 0.16    & 4       & &  $+$2.18 &  $+$0.10 & 0.08    & 0.15    & 3       \\
Ge~\textsc{i}  & 32 & $+$3.65 &  $+$0.62 &  $-$1.05 & 0.27    & 0.30    & 2       & &  $-$0.03 &  $-$1.20 & 0.25    & 0.28    & 3       \\
Rb~\textsc{i}  & 37 & $+$2.52 & $<+$1.70 & $<+$1.16 & \nodata & \nodata & 1       & & $<+$1.60 & $<+$1.56 & \nodata & \nodata & 1       \\
Sr~\textsc{ii} & 38 & $+$2.87 &  $+$0.98 &  $+$0.04 & 0.18    & 0.24    & 1       & &  $-$0.18 &  $-$0.59 & 0.09    & 0.19    & 2       \\
Y~\textsc{ii}  & 39 & $+$2.21 &  $-$0.10 &  $-$0.38 & 0.10    & 0.19    & 8       & &  $-$0.89 &  $-$0.64 & 0.13    & 0.21    & 7       \\
Zr~\textsc{ii} & 40 & $+$2.58 &  $+$0.74 &  $+$0.09 & 0.17    & 0.24    & 25      & &  $+$0.00 &  $-$0.12 & 0.11    & 0.20    & 27      \\
Nb~\textsc{ii} & 41 & $+$1.46 &  $-$0.36 &  $+$0.11 & 0.12    & 0.20    & 4       & &  $-$0.88 &  $+$0.12 & 0.21    & 0.27    & 1       \\
Mo~\textsc{i}  & 42 & $+$1.88 &  $+$0.18 &  $+$0.28 & 0.20    & 0.23    & 1       & &  $-$0.67 &  $-$0.07 & 0.13    & 0.18    & 1       \\
Mo~\textsc{ii} & 42 & $+$1.88 & $<+$0.43 & $<+$0.48 & \nodata & \nodata & 1       & & $<+$0.11 & $<+$0.69 & \nodata & \nodata & 1       \\
Ru~\textsc{i}  & 44 & $+$1.75 &  $+$0.09 &  $+$0.32 & 0.12    & 0.17    & 3       & &  $-$0.63 &  $+$0.10 & 0.15    & 0.19    & 1       \\
Rh~\textsc{i}  & 45 & $+$1.06 &  $-$0.65 &  $+$0.27 & 0.36    & 0.38    & 3       & & $<-$0.61 & $<+$0.81 & \nodata & \nodata & 1       \\
Pd~\textsc{i}  & 46 & $+$1.65 &  $-$0.32 &  $+$0.01 & 0.12    & 0.17    & 3       & &  $-$1.24 &  $-$0.41 & 0.24    & 0.27    & 1       \\
Ag~\textsc{i}  & 47 & $+$1.20 &  $-$1.01 &  $-$0.23 & 0.36    & 0.38    & 1       & & $<-$1.11 & $<+$0.17 & \nodata & \nodata & 1       \\
Cd~\textsc{i}  & 48 & $+$1.71 &  \nodata &  \nodata & \nodata & \nodata & \nodata & &  $-$1.38 &  $-$0.61 & 0.27    & 0.30    & 1       \\
Sn~\textsc{i}  & 50 & $+$2.07 &  \nodata &  \nodata & \nodata & \nodata & \nodata & &  \nodata &  \nodata & \nodata & \nodata & \nodata \\
Te~\textsc{i}  & 52 & $+$2.18 &  \nodata &  \nodata & \nodata & \nodata & \nodata & &  $-$0.18 &  $+$0.12 & 0.22    & 0.25    & 1       \\
Ba~\textsc{ii} & 56 & $+$2.18 &  $-$0.03 &  $-$0.28 & 0.13    & 0.21    & 2       & &  $-$1.03 &  $-$0.75 & 0.08    & 0.18    & 3       \\
La~\textsc{ii} & 57 & $+$1.10 &  $-$0.90 &  $-$0.07 & 0.06    & 0.18    & 9       & &  $-$1.64 &  $-$0.28 & 0.10    & 0.19    & 4       \\
Ce~\textsc{ii} & 58 & $+$1.58 &  $-$0.49 &  $-$0.14 & 0.06    & 0.18    & 13      & &  $-$1.15 &  $-$0.27 & 0.08    & 0.18    & 3       \\
Pr~\textsc{ii} & 59 & $+$0.72 &  $-$1.08 &  $+$0.13 & 0.13    & 0.21    & 2       & &  \nodata &  \nodata & \nodata & \nodata & \nodata \\
Nd~\textsc{ii} & 60 & $+$1.42 &  $-$0.52 &  $-$0.01 & 0.10    & 0.19    & 23      & &  $-$1.23 &  $-$0.19 & 0.06    & 0.18    & 4       \\
Sm~\textsc{ii} & 62 & $+$0.96 &  $-$0.93 &  $+$0.04 & 0.10    & 0.19    & 13      & &  $-$1.50 &  $+$0.00 & 0.18    & 0.24    & 2       \\
Eu~\textsc{ii} & 63 & $+$0.52 &  $-$1.19 &  $+$0.22 & 0.11    & 0.20    & 3       & &  $-$1.96 &  $-$0.02 & 0.06    & 0.18    & 4       \\
Gd~\textsc{ii} & 64 & $+$1.07 &  $-$0.68 &  $+$0.18 & 0.10    & 0.19    & 9       & &  \nodata &  \nodata & \nodata & \nodata & \nodata \\
Tb~\textsc{ii} & 65 & $+$0.30 &  $-$1.61 &  $+$0.02 & 0.20    & 0.26    & 1       & &  \nodata &  \nodata & \nodata & \nodata & \nodata \\
Dy~\textsc{ii} & 66 & $+$1.10 &  $-$0.60 &  $+$0.23 & 0.08    & 0.18    & 8       & &  $-$1.35 &  $+$0.01 & 0.11    & 0.20    & 4       \\
Ho~\textsc{ii} & 67 & $+$0.48 &  $-$1.37 &  $+$0.08 & 0.36    & 0.40    & 1       & &  $-$1.71 &  $+$0.27 & 0.24    & 0.29    & 1       \\
Er~\textsc{ii} & 68 & $+$0.92 &  $-$0.78 &  $+$0.23 & 0.20    & 0.26    & 1       & &  $-$1.54 &  $+$0.00 & 0.09    & 0.19    & 3       \\
Tm~\textsc{ii} & 69 & $+$0.10 &  $-$1.60 &  $+$0.23 & 0.12    & 0.20    & 3       & &  $-$2.24 &  $+$0.12 & 0.13    & 0.21    & 1       \\
Yb~\textsc{ii} & 70 & $+$0.92 &  $-$0.93 &  $+$0.08 & 0.20    & 0.26    & 1       & &  $-$1.69 &  $-$0.15 & 0.13    & 0.21    & 1       \\
Lu~\textsc{ii} & 71 & $+$0.10 &  \nodata &  \nodata & \nodata & \nodata & \nodata & &  $-$1.64 &  $+$0.72 & 0.21    & 0.27    & 1       \\
Hf~\textsc{ii} & 72 & $+$0.85 &  $-$0.99 &  $+$0.09 & 0.23    & 0.28    & 2       & &  $-$1.07 &  $+$0.54 & 0.21    & 0.27    & 1       \\
Os~\textsc{i}  & 76 & $+$1.40 &  $-$0.38 &  $+$0.20 & 0.39    & 0.41    & 1       & & $<-$0.39 & $<+$0.69 & \nodata & \nodata & 1       \\
Os~\textsc{ii} & 76 & $+$1.40 &  $-$0.86 &  $-$0.33 & 0.39    & 0.42    & 1       & & $<-$1.19 & $<-$0.13 & \nodata & \nodata & 1       \\
Ir~\textsc{i}  & 77 & $+$1.38 &  $-$0.28 &  $+$0.32 & 0.18    & 0.22    & 2       & & $<-$0.29 & $<+$0.81 & \nodata & \nodata & 2       \\
Pt~\textsc{i}  & 78 & $+$1.62 &  $-$0.34 &  $+$0.02 & 0.39    & 0.41    & 1       & & $<-$1.19 & $<-$0.33 & \nodata & \nodata & 2       \\
Au~\textsc{i}  & 79 & $+$0.80 &  \nodata &  \nodata & \nodata & \nodata & \nodata & & $<-$0.49 & $<+$1.19 & \nodata & \nodata & 1       \\
Pb~\textsc{i}  & 82 & $+$2.04 &  $+$0.16 &  $+$0.10 & 0.23    & 0.26    & 2       & & $<-$0.09 & $<+$0.35 & \nodata & \nodata & 1       \\
Bi~\textsc{i}  & 83 & $+$0.65 & $<+$1.43 & $<+$2.76 & \nodata & \nodata & 1       & & $<+$0.91 & $<+$2.74 & \nodata & \nodata & 1       \\
Th~\textsc{ii} & 90 & $+$0.06 &  \nodata &  \nodata & \nodata & \nodata & \nodata & &  \nodata &  \nodata & \nodata & \nodata & \nodata \\
\enddata
\end{deluxetable*}

%% file: tab9-stub.tex
\begin{deluxetable*}{ccccccc}
\tablecaption{Line-by-line Iron Abundances
\label{irontab}}
\tablewidth{0pt}
\tabletypesize{\scriptsize}
\tablehead{
\colhead{} &
\colhead{} &
\colhead{} &
\colhead{HD~108317} &
\colhead{HD~122563} &
\colhead{HD~126238} &
\colhead{HD~128279} \\
\colhead{Wavelength} &
\colhead{E.P.} &
\colhead{$\log(gf)$} &
\colhead{$\log\epsilon$} &
\colhead{$\log\epsilon$} &
\colhead{$\log\epsilon$} &
\colhead{$\log\epsilon$} \\
\colhead{(\AA)} &
\colhead{(eV)} &
\colhead{} &
\colhead{} &
\colhead{} &
\colhead{} &
\colhead{} }
\startdata
\multicolumn{7}{c}{Fe~\textsc{i}} \\
\hline
2283.30  & 0.12   &   $-$2.22  &   5.31     & \nodata  & \nodata  & 5.34    \\
\enddata
\tablecomments{
The complete version of Table~\ref{fetab} is available online only.
A short version is shown here to demonstrate its form and content.
Wavelength-dependent corrections derived from Fe~\textsc{i} have been
applied to the $\log\epsilon$ entries in this table.
The uncorrected $\log\epsilon$ values may be obtained by
subtracting the values listed in the final column of Table~\ref{corrtab}.
All $\log(gf)$ values have been adopted from the
critical compilation of \citet{fuhr06}.
 }
\end{deluxetable*}

%% file: tab10.tex
\begin{deluxetable}{ccccc}
\tablecaption{Mean Fe~\textsc{i} Abundances Binned by Wavelength
\label{corrtab}}
\tablewidth{0pt}
\tabletypesize{\scriptsize}
\tablehead{
\colhead{Wavelength} &
\colhead{$\langle\log\epsilon\rangle$} &
\colhead{Std.\ Dev} &
\colhead{$N$} &
\colhead{Correction} \\
\colhead{Range (\AA)} &
\colhead{} &
\colhead{} &
\colhead{} &
\colhead{} }
\startdata
\multicolumn{5}{c}{HD~108317} \\
\hline
2280--3100 & 4.93 & 0.163 & 32  & $+$0.04     \\
3100--3647 & 4.81 & 0.093 & 107 & $+$0.16     \\
3647--4000 & 4.87 & 0.081 & 40  & $+$0.10     \\
4000--4400 & 4.92 & 0.074 & 27  & $+$0.05     \\
4400--6750 & 4.97 & 0.070 & 94  & $\equiv$0.0 \\
\hline\hline
\multicolumn{5}{c}{HD~122563} \\
\hline
2280--3100 & 4.39 & 0.252 & 12 & $+$0.05     \\
3100--3647 & 4.17 & 0.112 & 69 & $+$0.27     \\
3647--4000 & 4.26 & 0.090 & 27 & $+$0.18     \\
4000--4400 & 4.39 & 0.055 & 17 & $+$0.05     \\
4400--6750 & 4.44 & 0.075 & 86 & $\equiv$0.0 \\
\hline\hline
\multicolumn{5}{c}{HD~126238} \\
\hline
2280--3100 & 5.39 & 0.197 & 10 & $+$0.13     \\
3100--3647 & 5.28 & 0.117 & 76 & $+$0.24     \\
3647--4000 & 5.37 & 0.108 & 22 & $+$0.15     \\
4000--4400 & 5.46 & 0.066 & 13 & $+$0.06     \\
4400--6750 & 5.52 & 0.082 & 96 & $\equiv$0.0 \\
\hline\hline
\multicolumn{5}{c}{HD~128279} \\
\hline
2280--3100 & 4.91 & 0.172 & 28 & $+$0.11     \\
3100--3647 & 4.83 & 0.096 & 96 & $+$0.19     \\
3647--4000 & 4.91 & 0.070 & 32 & $+$0.11     \\
4000--4400 & 4.97 & 0.067 & 19 & $+$0.05     \\
4400--6750 & 5.02 & 0.068 & 91 & $\equiv$0.0 \\
\enddata
\end{deluxetable}

%% file: tab11.tex
\begin{deluxetable*}{ccccccc}
\tablecaption{Hyperfine Structure Line Component Pattern for 
$^{175}$Lu~\textsc{ii} 2615\,\AA\
\label{lutab}}
\tablewidth{0pt}
\tabletypesize{\scriptsize}
\tablehead{
\colhead{Wavenumber} &
\colhead{$\lambda_{\rm air}$} &
\colhead{F$_{\rm upper}$} &
\colhead{F$_{\rm lower}$} &
\colhead{Component Position} &
\colhead{Component Position} &
\colhead{Strength} \\
\colhead{(cm$^{-1}$)} &
\colhead{(\AA)} &
\colhead{} &
\colhead{} &
\colhead{(cm$^{-1}$)} &
\colhead{(\AA)} &
\colhead{}
}
\startdata
 38223.406 & 2615.4173 & 4.5 & 3.5 & $-$0.11034 & $+$0.007550 & 0.41667 \\
 38223.406 & 2615.4173 & 3.5 & 3.5 & $-$0.02055 & $+$0.001406 & 0.33333 \\
 38223.406 & 2615.4173 & 2.5 & 3.5 & $+$0.21129 & $-$0.014458 & 0.25000 \\
\enddata
\end{deluxetable*}

%% file: tab12.tex
\begin{deluxetable*}{cccccccc}
\tablecaption{Hyperfine Structure and Isotopic Line Component Patterns for 
Pb~\textsc{i} Lines
\label{pbtab}}
\tablewidth{0pt}
\tabletypesize{\scriptsize}
\tablehead{
\colhead{Wavenumber} &
\colhead{$\lambda_{\rm air}$} &
\colhead{F$_{\rm upper}$} &
\colhead{F$_{\rm lower}$} &
\colhead{Component Position} &
\colhead{Component Position} &
\colhead{Strength} &
\colhead{Isotope} \\
\colhead{(cm$^{-1}$)} &
\colhead{(\AA)} &
\colhead{} &
\colhead{} &
\colhead{(cm$^{-1}$)} &
\colhead{(\AA)} &
\colhead{} &
\colhead{}
}
\startdata
35287.2545 & 2833.05104 & 1.0 & 0.0 & $-$0.11038 & $+$0.008862  & 0.014 & 204 \\
35287.2545 & 2833.05104 & 1.0 & 0.0 & $-$0.04415 & $+$0.003545  & 0.241 & 206 \\
35287.2545 & 2833.05104 & 1.5 & 0.5 & $+$0.13053 & $-$0.010480  & 0.147 & 207 \\
35287.2545 & 2833.05104 & 0.5 & 0.5 & $-$0.30988 & $+$0.024880  & 0.074 & 207 \\
35287.2545 & 2833.05104 & 1.0 & 0.0 & $+$0.03012 & $-$0.002418  & 0.524 & 208 \\
27467.9949 & 3639.56335 & 1.0 & 1.0 & $-$0.12122 & $+$0.016062  & 0.014 & 204 \\
27467.9949 & 3639.56335 & 1.0 & 1.0 & $-$0.04856 & $+$0.006435  & 0.241 & 206 \\
27467.9949 & 3639.56335 & 1.5 & 0.5 & $+$0.04930 & $-$0.006533  & 0.025 & 207 \\
27467.9949 & 3639.56335 & 1.5 & 1.5 & $+$0.16886 & $-$0.022374  & 0.123 & 207 \\
27467.9949 & 3639.56335 & 0.5 & 0.5 & $-$0.39110 & $+$0.051824  & 0.049 & 207 \\
27467.9949 & 3639.56335 & 0.5 & 1.5 & $-$0.27155 & $+$0.035982  & 0.025 & 207 \\
27467.9949 & 3639.56335 & 1.0 & 1.0 & $+$0.03308 & $-$0.004383  & 0.524 & 208 \\
27140.6795 & 3683.45748 & 0.0 & 1.0 & $-$0.12332 & $+$0.016737  & 0.014 & 204 \\
27140.6795 & 3683.45748 & 0.0 & 1.0 & $-$0.04938 & $+$0.006702  & 0.241 & 206 \\
27140.6795 & 3683.45748 & 0.5 & 0.5 & $-$0.09792 & $+$0.013290  & 0.074 & 207 \\
27140.6795 & 3683.45748 & 0.5 & 1.5 & $+$0.02163 & $-$0.002935  & 0.147 & 207 \\
27140.6795 & 3683.45748 & 0.0 & 1.0 & $+$0.03369 & $-$0.004572  & 0.524 & 208 \\
24636.9301 & 4057.80118 & 1.0 & 2.0 & $-$0.12020 & $+$0.019798  & 0.014 & 204 \\
24636.9301 & 4057.80118 & 1.0 & 2.0 & $-$0.04815 & $+$0.007931  & 0.241 & 206 \\
24636.9301 & 4057.80118 & 1.5 & 2.5 & $+$0.04233 & $-$0.006973  & 0.133 & 207 \\
24636.9301 & 4057.80118 & 1.5 & 1.5 & $+$0.25922 & $-$0.042695  & 0.015 & 207 \\
24636.9301 & 4057.80118 & 0.5 & 1.5 & $-$0.18119 & $+$0.029843  & 0.074 & 207 \\
24636.9301 & 4057.80118 & 1.0 & 2.0 & $+$0.03283 & $-$0.005407  & 0.524 & 208 \\
\enddata
\end{deluxetable*}

%% file: ms.bbl
\begin{thebibliography}{}


\bibitem[Alonso et al.(1999a)]{alonso99a} Alonso, A., Arribas, S., \& 
Mart{\'{\i}}nez-Roger, C.\ 1999a, \aaps, 139, 335 

\bibitem[Alonso et al.(1999b)]{alonso99b} Alonso, A., Arribas, S., \& 
Mart{\'{\i}}nez-Roger, C.\ 1999b, \aaps, 140, 261 

\bibitem[Alvarez \& Plez(1998)]{alvarez98} Alvarez, R., \& Plez, B.\ 
1998, \aap, 330, 1109 

\bibitem[Anders(1971)]{anders71} Anders, E.\ 1971, \gca, 35, 516 

\bibitem[Anders \& Grevesse(1989)]{anders89} Anders, E., \& Grevesse, N.\ 
1989, \gca, 53, 197 

\bibitem[Aoki et al.(2005)]{aoki05} Aoki, W., Honda, S., 
Beers, T.~C., et al.\ 2005, \apj, 632, 611 

\bibitem[Argast et al.(2004)]{argast04} Argast, D., Samland, M., 
Thielemann, F.-K., \& Qian, Y.-Z.\ 2004, \aap, 416, 997 

\bibitem[Arlandini et al.(1999)]{arlandini99} Arlandini, C., 
K{\"a}ppeler, F., Wisshak, K., et al.\ 1999, \apj, 525, 886 

\bibitem[Asplund et al.(2009)]{asplund09} Asplund, M., Grevesse, N., 
Sauval, A.~J., \& Scott, P.\ 2009, \araa, 47, 481 

\bibitem[Barbuy et al.(2011)]{barbuy11} Barbuy, B., Spite, M., Hill, V., 
et al.\ 2011, \aap, 534, A60 

\bibitem[Barklem et al.(2000)]{barklem00} Barklem, P.~S., Piskunov, N., \& 
O'Mara, B.~J.\ 2000, \aaps, 142, 467 

\bibitem[Barklem et al.(2005)]{barklem05} Barklem, P.~S., Christlieb, N., 
Beers, T.~C., et al.\ 2005, \aap, 439, 129 

\bibitem[Barklem \& Aspelund-Johansson(2005)]{barklem05b} Barklem, P.~S., \& 
Aspelund-Johansson, J.\ 2005, \aap, 435, 373 

\bibitem[Bernstein et al.(2003)]{bernstein03} Bernstein, R., 
Shectman, S.~A., Gunnels, S.~M., Mochnacki, S., 
\& Athey, A.~E.\ 2003, \procspie, 4841, 1694 

\bibitem[Bielski(1975)]{bielski75} Bielski, A.\ 1975, \jqsrt, 15, 463 

\bibitem[Bi{\'e}mont \& Godefroid(1980)]{biemont80} Bi{\'e}mont, E., \& 
Godefroid, M.\ 1980, \aap, 84, 361 

\bibitem[Bi{\'e}mont et al.(1999)]{biemont99} Bi{\'e}mont, E., Lynga, C., 
Li, Z.~S., et al.\ 1999, \mnras, 303, 721 

\bibitem[Bi{\'e}mont et al.(2000)]{biemont00} Bi{\'e}mont, E., 
Garnir, H.~P., Palmeri, P., Li, Z.~S., 
\& Svanberg, S.\ 2000, \mnras, 312, 116 

\bibitem[Bi{\'e}mont et al.(2011)]{biemont11} Bi{\'e}mont, 
{\'E}., Blagoev, K., Engstr{\"o}m, L., et al.\ 2011, \mnras, 414, 3350 

\bibitem[Bisterzo et al.(2010)]{bisterzo10} Bisterzo, S., Gallino, 
R., Straniero, O., Cristallo, S., K\"{a}ppeler, F.\ 2010, \mnras, 404, 1529 

\bibitem[Bisterzo et al.(2011)]{bisterzo11} Bisterzo, S., Gallino, R., 
Straniero, O., Cristallo, S., \& K\"{a}ppeler, F.\ 2011, \mnras, 418, 284

\bibitem[Bohlin et al.(1978)]{bohlin78} Bohlin, R.~C., Savage, 
B.~D., \& Drake, J.~F.\ 1978, \apj, 224, 132 

\bibitem[Bonifacio et al.(2000)]{bonifacio00} Bonifacio, P., Monai, 
S., \& Beers, T.~C.\ 2000, \aj, 120, 2065 

\bibitem[Bouazza et al.(2000)]{bouazza00} Bouazza, S., Gough, D.~S., 
Hannaford, P., Lowe, R.~M., \& Wilson, M.\ 2000, Phys.\ Rev.\ A, 63, 012516

\bibitem[B\"{o}hlke et al.(2005)]{bohlke05} B\"{o}hlke, J.~K., 
de Laeter, J.~R., De Bi\'{e}vre, P., Hidaka, H., Peiser, H.~S., et al.\
2005, J.\ Phys.\ Chem.\ Ref.\ Data, 34, 57

\bibitem[Burris et al.(2000)]{burris00} Burris, D.~L., 
Pilachowski, C.~A., Armandroff, T.~E., et al.\ 2000, \apj, 544, 302 

\bibitem[Busso et al.(1999)]{busso99} Busso, M., Gallino, R., \& 
Wasserburg, G.~J.\ 1999, \araa, 37, 239 

\bibitem[Cameron(1973)]{cameron73} Cameron, A.~G.~W.\ 1973, \ssr, 15, 121 

\bibitem[Cameron(1982)]{cameron82} Cameron, A.~G.~W.\ 1982, \apss, 82, 123 

\bibitem[Cardelli et al.(1989)]{cardelli89} Cardelli, J.~A., 
Clayton, G.~C., \& Mathis, J.~S.\ 1989, \apj, 345, 245 

\bibitem[Castelli \& Kurucz(2004)]{castelli04} Castelli, F., \& Kurucz, R.~L.\
2004, in IAU Symp.\ 210, Modelling of Stellar Atmospheres, ed.\ N.\ Piskunov
et al.\ (Cambridge: Cambridge Univ.\ Press), A20

\bibitem[Cohen et al.(2008)]{cohen08} Cohen, J.~G., Christlieb, 
N., McWilliam, A., et al.\ 2008, \apj, 672, 320 

\bibitem[Cowan et al.(1995)]{cowan95} Cowan, J.~J., Burris, 
D.~L., Sneden, C., McWilliam, A., \& Preston, G.~W.\ 1995, \apjl, 439, L51 

\bibitem[Cowan et al.(1996)]{cowan96} Cowan, J.~J., Sneden, C., 
Truran, J.~W., \& Burris, D.~L.\ 1996, \apjl, 460, L115 

\bibitem[Cowan et al.(2002)]{cowan02} Cowan, J.~J., Sneden, C., 
Burles, S., et al.\ 2002, \apj, 572, 861 

\bibitem[Cowan et al.(2005)]{cowan05} Cowan, J.~J., Sneden, C., 
Beers, T.~C., et al.\ 2005, \apj, 627, 238 

\bibitem[Creevey et al.(2012)]{creevey12} Creevey, O.~L., Th{\'e}venin, F., 
Boyajian, T.~S., et al.\ 2012, \aap, 545, A17 

\bibitem[D\"{a}ppen et al.(1987)]{dappen87} D\"{a}ppen, W., Anderson, 
L., \& Mihalas, D.\ 1987, \apj, 319, 195 

\bibitem[Den Hartog et al.(2003)]{denhartog03} Den Hartog, E.~A., 
Lawler, J.~E., Sneden, C., \& Cowan, J.~J.\ 2003, \apjs, 148, 543 

\bibitem[Den Hartog et al.(2005)]{denhartog05} Den Hartog, E.~A., 
Herd, M.~T., Lawler, J.~E., et al.\ 2005, \apj, 619, 639 

\bibitem[Den Hartog et al.(2006)]{denhartog06} Den Hartog, E.~A., 
Lawler, J.~E., Sneden, C., \& Cowan, J.~J.\ 2006, \apjs, 167, 292 

\bibitem[Duquette \& Lawler(1985)]{duquette85} Duquette, D.~W., \& 
Lawler, J.~E.\ 1985, Journal of the Optical Society of America B 
Optical Physics, 1, 1948 

\bibitem[Farouqi et al.(2010)]{farouqi10} Farouqi, K., Kratz, 
K.-L., Pfeiffer, B., et al.\ 2010, \apj, 712, 1359 

\bibitem[Ferlet et al.(1985)]{ferlet85} Ferlet, R., 
Vidal-Madjar, A., \& Gry, C.\ 1985, \apj, 298, 838 

\bibitem[Fischer et al.(2010)]{fischer10} Fischer, T., Whitehouse, S.~C., 
Mezzacappa, A., Thielemann, F.-K., \& Liebend{\"o}rfer, M.\ 2010, \aap, 
517, A80 

\bibitem[Fivet et al.(2006)]{fivet06} Fivet, V., Quinet, P., 
Bi{\'e}mont, {\'E}., 
\& Xu, H.~L.\ 2006, Journal of Physics B Atomic Molecular Physics, 39, 3587 

\bibitem[Fran{\c c}ois et al.(2007)]{francois07} Fran{\c c}ois, P., 
Depagne, E., Hill, V., et al.\ 2007, \aap, 476, 935 

\bibitem[Frebel et al.(2005)]{frebel05} Frebel, A., Aoki, W., 
Christlieb, N., et al.\ 2005, \nat, 434, 871 

\bibitem[Frebel et al.(2007)]{frebel07} Frebel, A., Christlieb, 
N., Norris, J.~E., et al.\ 2007, \apjl, 660, L117 

\bibitem[Frebel et al.(2010a)]{frebel10} Frebel, A., Kirby, 
E.~N., \& Simon, J.~D.\ 2010a, \nat, 464, 72 

\bibitem[Frebel et al.(2010b)]{frebel10b} Frebel, A., Simon, 
J.~D., Geha, M., \& Willman, B.\ 2010b, \apj, 708, 560 

\bibitem[Freiburghaus et al.(1999)]{freiburghaus99} Freiburghaus, C., 
Rosswog, S., \& Thielemann, F.-K.\ 1999, \apjl, 525, L121 

\bibitem[Fuhr \& Wiese(2006)]{fuhr06} Fuhr, J.~R., \& Wiese, W.~L.\ 
2006, Journal of Physical and Chemical Reference Data, 35, 1669 

\bibitem[Fuhr \& Wiese(2009)]{fuhr09} Fuhr, J.~R.\ \& Wiese, W.~L.\ 2009,
Atomic Transition Probabilities, published in the CRC Handbook of Chemistry 
and Physics, 90th Edition, ed.\ Lide, D.~R., CRC Press, Inc., 
Boca Raton, FL, 10

\bibitem[Fulbright et al.(2004)]{fulbright04} Fulbright, J.~P., 
Rich, R.~M., \& Castro, S.\ 2004, \apj, 612, 447 

\bibitem[Gallino et al.(1998)]{gallino98} Gallino, R., Arlandini, 
C., Busso, M., et al.\ 1998, \apj, 497, 388 

\bibitem[Goriely et al.(2011)]{goriely11} Goriely, S., Bauswein, 
A., \& Janka, H.-T.\ 2011, \apjl, 738, L32 

\bibitem[Hansen et al.(2011)]{hansen11b} Hansen, T., Andersen, 
J., Nordstr{\"o}m, B., Buchhave, L.~A., 
\& Beers, T.~C.\ 2011, \apjl, 743, L1 

\bibitem[Hansen \& Primas(2011)]{hansen11} Hansen, C.~J., \& Primas, F.\ 
2011, \aap, 525, L5 

\bibitem[Hansen et al.(2012)]{hansen12} Hansen, C.~J., Primas, F., 
Hartman, H., et al.\ 2012, \aap, 545, A31

\bibitem[Hill et al.(2002)]{hill02} Hill, V., Plez, B., Cayrel, R., 
et al.\ 2002, \aap, 387, 560 

\bibitem[Hinkle et al.(2000)]{hinkle00} Hinkle, K., Wallace, L., 
Valenti, J., \& Harmer, D.\ 2000, Visible and Near Infrared Atlas of the 
Arcturus Spectrum 3727-9300~A, ed.\ K.\ Hinkle, L.\ Wallace, J.\ Valenti, 
D.\ Harmer.\ (San Francisco: ASP)

\bibitem[Hinkle et al.(2005)]{hinkle05} Hinkle, K., Wallace, L., 
Valenti, J., \& Ayres, T.\ 2005, Ultraviolet Atlas of the Arcturus Spectrum, 
1150-3800~A, ed.\ K.\ Hinkle, L.\ Wallace, J.\ Valenti, T.\ Ayres.\
 (San Francisco: ASP) 

\bibitem[Hollek et al.(2011)]{hollek11} Hollek, J.~K., Frebel, 
A., Roederer, I.~U., et al.\ 2011, \apj, 742, 54 

\bibitem[Honda et al.(2004)]{honda04} Honda, S., Aoki, W., 
Kajino, T., et al.\ 2004, \apj, 607, 474 

\bibitem[Honda et al.(2006)]{honda06} Honda, S., Aoki, W., 
Ishimaru, Y., Wanajo, S., \& Ryan, S.~G.\ 2006, \apj, 643, 1180 

\bibitem[Honda et al.(2007)]{honda07} Honda, S., Aoki, W., 
Ishimaru, Y., \& Wanajo, S.\ 2007, \apj, 666, 1189 

\bibitem[Horowitz \& Li(1999)]{horowitz99} Horowitz, C.~J., \& Li, G.\ 1999, 
Physical Review Letters, 82, 5198 

\bibitem[Hubeny et al.(1994)]{hubeny94} Hubeny, I., Hummer, D.~G., \& 
Lanz, T.\ 1994, \aap, 282, 151 

\bibitem[H{\"u}depohl et al.(2010)]{hudepohl10} H{\"u}depohl, L., 
M{\"u}ller, B., Janka, H.-T., Marek, A., 
\& Raffelt, G.~G.\ 2010, Physical Review Letters, 104, 251101 

\bibitem[Hummer \& Mihalas(1988)]{hummer88} Hummer, D.~G., \& 
Mihalas, D.\ 1988, \apj, 331, 794 

\bibitem[Ivans et al.(2006)]{ivans06} Ivans, I.~I., Simmerer, 
J., Sneden, C., et al.\ 2006, \apj, 645, 613 

\bibitem[Ivarsson et al.(2004)]{ivarsson04} Ivarsson, S., Wahlgren, G.~M., 
Dai, Z., Lundberg, H., \& Leckrone, D.~S.\ 2004, \aap, 425, 353 

\bibitem[Janka et al.(2008)]{janka08} Janka, H.-T., M{\"u}ller, B., 
Kitaura, F.~S., \& Buras, R.\ 2008, \aap, 485, 199 

\bibitem[Johnson \& Bolte(2002)]{johnson02} Johnson, J.~A., \& Bolte, M.\ 
2002, \apj, 579, 616 

\bibitem[K\"{a}ppeler et al.(1989)]{kappeler89} K\"{a}ppeler, F., Beer, 
H., \& Wisshak, K.\ 1989, Reports on Progress in Physics, 52, 945 

\bibitem[Karzas \& Latter(1961)]{karzas61} Karzas, W.~J., \& Latter, R.\ 
1961, \apjs, 6, 167 

\bibitem[Kelson(2003)]{kelson03} Kelson, D.~D.\ 2003, \pasp, 115, 688 

\bibitem[Kimble et al.(1998)]{kimble98} Kimble, R.~A., Woodgate, 
B.~E., Bowers, C.~W., et al.\ 1998, \apjl, 492, L83 

\bibitem[Kling \& Kock(1999)]{kling99} Kling, R., \& Kock, M.\ 1999, 
\jqsrt, 62, 129 

\bibitem[Koch et al.(2008)]{koch08} Koch, A., McWilliam, A., 
Grebel, E.~K., Zucker, D.~B., \& Belokurov, V.\ 2008, \apjl, 688, L13 

\bibitem[Kratz et al.(2007)]{kratz07} Kratz, K.-L., Farouqi, 
K., Pfeiffer, B., et al.\ 2007, \apj, 662, 39 

\bibitem[Kurucz(1970)]{kurucz70} Kurucz, R.~L.\ 1970, SAO Special 
Report 309, ``Atlas: A computer program for calculating model stellar 
atmospheres,'' Cambridge: Smithsonian Astrophysical Observatory

\bibitem[Kurucz \& Bell(1995)]{kurucz95} Kurucz, R.~L., \& Bell, B.\ 1995, 
Kurucz CD-ROM, Cambridge, MA: Smithsonian Astrophysical Observatory

\bibitem[Kwiatkowski et al.(1982)]{kwiatkowski82} Kwiatkowski, M., 
Zimmermann, P., Bi{\'e}mont, E., \& Grevesse, N.\ 1982, \aap, 112, 337 

\bibitem[Lai et al.(2008)]{lai08} Lai, D.~K., Bolte, M., 
Johnson, J.~A., et al.\ 2008, \apj, 681, 1524 

\bibitem[Lawler et al.(2001a)]{lawler01a} Lawler, J.~E., 
Bonvallet, G., \& Sneden, C.\ 2001a, \apj, 556, 452 

\bibitem[Lawler et al.(2001b)]{lawler01b} Lawler, J.~E., 
Wickliffe, M.~E., den Hartog, E.~A., \& Sneden, C.\ 2001b, \apj, 563, 1075 

\bibitem[Lawler et al.(2001c)]{lawler01c} Lawler, J.~E., 
Wickliffe, M.~E., Cowley, C.~R., \& Sneden, C.\ 2001c, \apjs, 137, 341 

\bibitem[Lawler et al.(2004)]{lawler04} Lawler, J.~E., Sneden, C., \& 
Cowan, J.~J.\ 2004, \apj, 604, 850 

\bibitem[Lawler et al.(2006)]{lawler06} Lawler, J.~E., Den 
Hartog, E.~A., Sneden, C., \& Cowan, J.~J.\ 2006, \apjs, 162, 227 

\bibitem[Lawler et al.(2007)]{lawler07} Lawler, J.~E., Den 
Hartog, E.~A., Labby, Z.~E., Sneden, C., Cowan, J.~J., 
\& Ivans, I.~I.\ 2007, \apjs, 169, 120 

\bibitem[Lawler et al.(2008)]{lawler08} Lawler, J.~E., Sneden, 
C., Cowan, J.~J., Wyart, J.-F., Ivans, I.~I., Sobeck, J.~S., Stockett, 
M.~H., \& Den Hartog, E.~A.\ 2008, \apjs, 178, 71 

\bibitem[Lawler et al.(2009)]{lawler09} Lawler, J.~E., Sneden, 
C., Cowan, J.~J., Ivans, I.~I., \& Den Hartog, E.~A.\ 2009, \apjs, 182, 51 

\bibitem[Ljung et al.(2006)]{ljung06} Ljung, G., Nilsson, H., Asplund, M., 
\& Johansson, S.\ 2006, \aap, 456, 1181 

\bibitem[Leckrone et al.(1999)]{leckrone99} Leckrone, D.~S., 
Proffitt, C.~R., Wahlgren, G.~M., Johansson, S.~G., 
\& Brage, T.\ 1999, \aj, 117, 1454 

\bibitem[Li et al.(2007)]{li07} Li, R., Chatelain, R., Holt, 
R.~A., Rehse, S.~J., Rosner, S.~D., 
\& Scholl, T.~J.\ 2007, \physscr, 76, 577 

\bibitem[Lugaro et al.(2012)]{lugaro12} Lugaro, M., Karakas, 
A.~I., Stancliffe, R.~J., \& Rijs, C.\ 2012, \apj, 747, 2

\bibitem[Malcheva et al.(2006)]{malcheva06} Malcheva, G., Blagoev, 
K., Mayo, R., et al.\ 2006, \mnras, 367, 754 

\bibitem[Mashonkina et al.(2012)]{mashonkina12} Mashonkina, L., Ryabtsev, A., 
\& Frebel, A.\ 2012, \aap, 540, A98 

\bibitem[Mathews et al.(1992)]{mathews92} Mathews, G.~J., Bazan, 
G., \& Cowan, J.~J.\ 1992, \apj, 391, 719 

\bibitem[McClure(1984)]{mcclure84} McClure, R.~D.\ 1984, \apjl, 280, L31 

\bibitem[McClure \& Woodsworth(1990)]{mcclure90} McClure, R.~D., \& 
Woodsworth, A.~W.\ 1990, \apj, 352, 709 

\bibitem[McWilliam et al.(1995)]{mcwilliam95} McWilliam, A., 
Preston, G.~W., Sneden, C., \& Searle, L.\ 1995, \aj, 109, 2757 

\bibitem[McWilliam(1998)]{mcwilliam98} McWilliam, A.\ 1998, \aj, 115, 1640 

\bibitem[Migdalek \& Baylis(1987)]{migdalek87} Migdalek, J., \& 
Baylis, W.~E.\ 1987, Canadian Journal of Physics, 65, 1612 

\bibitem[Montes et al.(2007)]{montes07} Montes, F., Beers, 
T.~C., Cowan, J., et al.\ 2007, \apj, 671, 1685 

\bibitem[Morton(2000)]{morton00} Morton, D.~C.\ 2000, \apjs, 130, 403 

\bibitem[Nilsson et al.(2002)]{nilsson02} Nilsson, H., Zhang, Z.~G., 
Lundberg, H., Johansson, S., \& Nordstr{\"o}m, B.\ 2002, \aap, 382, 368 

\bibitem[Nilsson \& Ivarsson(2008)]{nilsson08} Nilsson, H., \& Ivarsson, S.\ 
2008, \aap, 492, 609 

\bibitem[Nilsson et al.(2010)]{nilsson10} Nilsson, H., Hartman, H., 
Engstr{\"o}m, L., et al.\ 2010, \aap, 511, A16 

\bibitem[Peck \& Reeder(1972)]{peck72} Peck, E.~R., \& Reeder, K.\ 1972, 
Journal of the Optical Society of America (1917-1983), 62, 958 

\bibitem[Perryman et al.(1997)]{perryman97} Perryman, M.~A.~C., 
Lindegren, L., Kovalevsky, J., et al.\ 1997, \aap, 323, L49 

\bibitem[Peterson(2011)]{peterson11} Peterson, R.~C.\ 2011, \apj, 742, 21 

\bibitem[Qian \& Wasserburg(2007)]{qian07} Qian, Y.-Z., \& 
Wasserburg, G.~J.\ 2007, \physrep, 442, 237 

\bibitem[Qian \& Wasserburg(2008)]{qian08} Qian, Y.-Z., \& 
Wasserburg, G.~J.\ 2008, \apj, 687, 272 

\bibitem[Quinet et al.(2006)]{quinet06} Quinet, P., Palmeri, P., 
Bi{\'e}mont, {\'E}., et al.\ 2006, \aap, 448, 1207 

\bibitem[Raiteri et al.(1993)]{raiteri93} Raiteri, C.-M., Gallino, R., 
Busso, M., Neuberger, D., \& K\"{a}ppeler, F.\ 1993, \apj, 419, 207

\bibitem[Ralchenko et al.(2011)]{ralchenko11}
Ralchenko, Yu., Kramida, A., Reader, J., et al.\ 2011, NIST Atomic Spectra
Database, version~4.1, available online: http://physics.nist.gov/asd

\bibitem[Roberts et al.(2010)]{roberts10} Roberts, L.~F., 
Woosley, S.~E., \& Hoffman, R.~D.\ 2010, \apj, 722, 954 

\bibitem[Roederer et al.(2008)]{roederer08} Roederer, I.~U., 
Lawler, J.~E., Sneden, C., et al.\ 2008, \apj, 675, 723 

\bibitem[Roederer et al.(2009)]{roederer09} Roederer, I.~U., 
Kratz, K.-L., Frebel, A., et al.\ 2009, \apj, 698, 1963 

\bibitem[Roederer et al.(2010a)]{roederer10a} Roederer, I.~U., 
Sneden, C., Thompson, I.~B., Preston, G.~W., 
\& Shectman, S.~A.\ 2010a, \apj, 711, 573 

\bibitem[Roederer et al.(2010b)]{roederer10b} Roederer, I.~U., 
Sneden, C., Lawler, J.~E., \& Cowan, J.~J.\ 2010b, \apjl, 714, L123 

\bibitem[Roederer et al.(2010c)]{roederer10c} Roederer, I.~U., 
Cowan, J.~J., Karakas, A.~I., et al.\ 2010c, \apj, 724, 975 

\bibitem[Roederer(2011)]{roederer11} Roederer, I.~U.\ 2011, \apjl, 732, L17 

\bibitem[Roederer et al.(2012)]{roederer12a} Roederer, I.~U., Lawler, J.~E.,
Cowan, J.~J., et al.\ 2012, \apjl, 747, L8

\bibitem[Roederer \& Lawler(2012)]{roederer12b} Roederer, I.~U., \& Lawler, 
J.~E.\ 2012, \apj, 750, 76

\bibitem[Roederer(2012a)]{roederer12c} Roederer, I.~U.\
2012a, \apj, 756, 36

\bibitem[Roederer(2012b)]{roederer12d} Roederer, I.~U.\
2012b, \aj, submitted

\bibitem[Ryan et al.(1996)]{ryan96} Ryan, S.~G., Norris, 
J.~E., \& Beers, T.~C.\ 1996, \apj, 471, 254 

\bibitem[Schlegel et al.(1998)]{schlegel98} Schlegel, D.~J., 
Finkbeiner, D.~P., \& Davis, M.\ 1998, \apj, 500, 525 

\bibitem[Seeger et al.(1965)]{seeger65} Seeger, P.~A., Fowler, 
W.~A., \& Clayton, D.~D.\ 1965, \apjs, 11, 121 

\bibitem[Sikstr{\"o}m et al.(2001)]{sikstrom01} Sikstr{\"o}m, 
C.~M., Pihlemark, H., Nilsson, H., et al.\ 2001, Journal of Physics B 
Atomic Molecular Physics, 34, 477 

\bibitem[Skrutskie et al.(2006)]{skrutskie06} Skrutskie, M.~F., 
Cutri, R.~M., Stiening, R., et al.\ 2006, \aj, 131, 1163 

\bibitem[Sneden(1973)]{sneden73} Sneden, C.~A.\ 1973, 
Ph.D.~Thesis, Univ.\ of Texas at Austin

\bibitem[Sneden \& Parthasarathy(1983)]{sneden83} Sneden, C., \& 
Parthasarathy, M.\ 1983, \apj, 267, 757 

\bibitem[Sneden et al.(1998)]{sneden98} Sneden, C., Cowan, 
J.~J., Burris, D.~L., \& Truran, J.~W.\ 1998, \apj, 496, 235 

\bibitem[Sneden et al.(2000)]{sneden00} Sneden, C., Cowan, 
J.~J., Ivans, I.~I., et al.\ 2000, \apjl, 533, L139 

\bibitem[Sneden et al.(2003)]{sneden03} Sneden, C., Cowan, 
J.~J., Lawler, J.~E., et al.\ 2003, \apj, 591, 936 

\bibitem[Sneden et al.(2008)]{sneden08} Sneden, C., Cowan, J.~J., \&
Gallino, R.\ 2008, \araa, 46, 241

\bibitem[Sneden et al.(2009)]{sneden09} Sneden, C., Lawler, J.~E., 
Cowan, J.~J., Ivans, I.~I., \& Den Hartog, E.~A.\ 2009, \apjs, 182, 80 

\bibitem[Sobeck et al.(2011)]{sobeck11} Sobeck, J.~S., Kraft, 
R.~P., Sneden, C., et al.\ 2011, \aj, 141, 175 

\bibitem[Thompson et al.(1983)]{thompson83} Thompson, R.~C., Anselment, M., 
Bekk, K., G\"{o}ring, S., Hanser, A., et al.\ 1983, J.\ Phys.\ G: Nucl.\ 
Phys., 9, 443

\bibitem[Travaglio et al.(2004)]{travaglio04} Travaglio, C., 
Gallino, R., Arnone, E., et al.\ 2004, \apj, 601, 864 

\bibitem[Truran et al.(2002)]{truran02} Truran, J.~W., Cowan, 
J.~J., Pilachowski, C.~A., \& Sneden, C.\ 2002, \pasp, 114, 1293 

\bibitem[Uns\"{o}ld(1955)]{unsold55} Uns\"{o}ld, A., Physik der
Sternatmosph\"{a}ren, Springer-Verlag, Berlin, 1955, p.\ 332--333

\bibitem[van Leeuwen(2007)]{vanleeuwen07} van Leeuwen, F.\ 2007, \aap, 474, 653 

\bibitem[Vogt et al.(1994)]{vogt94} Vogt, S.~S., Allen, S.~L., 
Bigelow, B.~C., et al.\ 1994, \procspie, 2198, 362 

\bibitem[Wanajo et al.(2003)]{wanajo03} Wanajo, S., Tamamura, 
M., Itoh, N., et al.\ 2003, \apj, 593, 968 

\bibitem[Wanajo \& Ishimaru(2006)]{wanajo06} Wanajo, S., \& 
Ishimaru, Y.\ 2006, Nuclear Physics A, 777, 676 

\bibitem[Wasserburg et al.(1996)]{wasserburg96} Wasserburg, G.~J., 
Busso, M., \& Gallino, R.\ 1996, \apjl, 466, L109 

\bibitem[Whaling \& Brault(1988)]{whaling88} Whaling, W., \& Brault, J.~W.\ 
1988, \physscr, 38, 707 

\bibitem[Wheeler et al.(1998)]{wheeler98} Wheeler, J.~C., Cowan, 
J.~J., \& Hillebrandt, W.\ 1998, \apjl, 493, L101 

\bibitem[Wickliffe et al.(1994)]{wickliffe94} Wickliffe, M.~E., 
Salih, S., \& Lawler, J.~E.\ 1994, \jqsrt, 51, 545 

\bibitem[Wickliffe \& Lawler(1997)]{wickliffe97} Wickliffe, M.~E., \& 
Lawler, J.~E.\ 1997, Journal of the Optical Society of America B 
Optical Physics, 14, 737 

\bibitem[Wickliffe et al.(2000)]{wickliffe00} Wickliffe, M.~E., 
Lawler, J.~E., \& Nave, G.\ 2000, J.\ Quant.\ Spectr.\ Radiat.\ Trans., 
66, 363 

\bibitem[Wood \& Andrew(1968)]{wood68} Wood, D.~R., \& Andrew, K.~L.\ 
1968, Journal of the Optical Society of America (1917-1983), 58, 818 

\bibitem[Woodgate et al.(1998)]{woodgate98} Woodgate, B.~E., 
Kimble, R.~A., Bowers, C.~W., et al.\ 1998, \pasp, 110, 1183 

\bibitem[Woosley et al.(1994)]{woosley94} Woosley, S.~E., Wilson, 
J.~R., Mathews, G.~J., Hoffman, R.~D., 
\& Meyer, B.~S.\ 1994, \apj, 433, 229 

\bibitem[Xu et al.(2006)]{xu06} Xu, H.~L., Sun, Z.~W., Dai, Z.~W., et al.\ 
2006, \aap, 452, 357 

\bibitem[Xu et al.(2007)]{xu07} Xu, H.~L., Svanberg, S., 
Quinet, P., Palmeri, P., \& Bi{\'e}mont, {\'E}.\ 2007, \jqsrt, 104, 52 

\bibitem[Zheng et al.(2011)]{zheng11} Zheng, W., Proffit, C., \&
Sahnow, D.\ 2011, Instrument Science Report STIS 2011-03, 1

\end{thebibliography}
